\documentclass[prd,a4paper,nofootinbib,floatfix,onecollumn,notitlepage]{revtex4-1}
\usepackage{amsmath}
\usepackage{graphicx}

\begin{document}

\title{Spherical steady accretion flows --- dependence on the cosmological constant, exact isothermal solutions and applications to cosmology}
\author{Patryk Mach, Edward Malec, Janusz Karkowski}
\affiliation{M.~Smoluchowski Institute of Physics, Jagiellonian University, Reymonta 4, 30-059 Krak\'{o}w}

\begin{abstract}
We investigate spherical, isothermal and polytropic steady accretion models in the presence of the cosmological constant. Exact solutions are found for three classes of isothermal fluids, assuming the test gas approximation. The cosmological constant damps the mass accretion rate and --- above certain limit --- completely stops the steady accretion onto  black holes.  A ``homoclinic-type'' accretion flow of polytropic gas has been discovered in AdS spacetimes in the test-gas limit. These results can have cosmological connotation, through the Einstein--Straus vacuole model of embedding local structures into Friedman--Lemaitre--Robertson--Walker spacetimes. In particular one infers that steady accretion would not exist in the late phases of the Penrose's scenario of the evolution of the Universe, known as the Weyl curvature hypothesis.
 
 \end{abstract}

\maketitle

\section{Introduction}

There are still serious reasons, in our opinion, to investigate Bondi-type accretion models \cite{bondi}. Firstly, they are still not yet fully explored in the general-relativistic context, despite a significant work that has been done in the last decades (see \cite{michel, shapiro, malec_1999, Mach_stability2, Mach2008, Mach_stability1, Rembiasz2010, karkowski_malec_2013} and many of forthcoming references). Secondly, and even more importantly, when put in the framework of cosmological spacetimes they  can be thought of  being simple toy models that shed light on the possible influence of the cosmological constant $\Lambda$ onto the formation of material structures. We obviously do not mean that these models can reproduce the arrangement of matter in, say, superclusters of galaxies; we expect, however, that if the cosmological constant damps Bondi accretion, then realistic models should reveal less robust structure formation. Thirdly --- and this essentially constitutes a reversal of the former argument --- if accretion is $\Lambda$-dependent, then the natural question is to ask under what circumstances it is  possible to estimate the value of the cosmological constant just by a quasilocal analysis? 

In a recent paper \cite{karkowski_malec_2013} two of us investigated stationary, spherically symmetric accretion flows of self-gravitating polytropic fluids in cosmological spacetimes. The physical picture discussed in \cite{karkowski_malec_2013} was as follows. We considered a spherical cloud of self-gravitating polytropic gas, accreting steadily on a central black hole. The outer areal radius of the cloud, its total mass and the mass of the black hole were fixed. We also fixed the parameters of the gas: the polytropic exponent and the local speed of sound at the outer boundary. The numerical data presented in \cite{karkowski_malec_2013} suggested that the mass accretion rate of transonic solutions decreases with the increasing modulus of the cosmological constant, irrespective of its sign, provided that the mass of gas is kept constant.   

In this paper we investigate spherically symmetric, steady flows of test-gases, for polytropic and isothermal equations of state. The emerging picture turns out to be more complex than it could be inferred form the previous work. In particular, the relation between the accretion rate and the cosmological constant depends on the assumed details of the model.
 
Section II of this paper specifies the de Sitter and anti-de Sitter (AdS) spacetimes and give equations that are relevant for the description of steady accretion. Section III shows the existence of ``homoclinic'' solutions, i.e., solutions for which the graph of the square of the radial velocity versus the areal radius forms a closed loop joining the same critical point. This phenomenon  happens only for polytropic equations of state and negative cosmological constant. A similar behavior was observed in approximate analysis of accretion models with low angular momentum \cite{tapas_czerny}. To our knowledge, it is the first time, where such solutions are obtained strictly and in spherical symmetry.

Section IV discusses transonic flows of isothermal fluids in the test gas limit. We find exact solutions and investigate the dependence of the mass accretion rate on $\Lambda$. In most cases, when keeping fixed the asymptotic mass density, the mass accretion rate decreases with the increase of $\Lambda$. The exception is the stiff equation of state  $p = e$, where $p$ is the pressure and $e$ denotes the energy density. Thence the baryonic mass accretion rate achieves a maximum at some $\Lambda <0$. 

In Section V we consider accretion in suitably prepared --- Swiss-cheese vacuolized \cite{Einstein1} --- Friedman--Lemaitre--Roberston--Walker (FLRW) models. Accreting system are put in the interiors of the vacuoles, in a way that does keep the overall FLRW geometry --- that means that the total mass of vacuoles is fixed. Then we study the dependence of accretion onto cosmological constant. It appears that polytropic and isothermal fluids behave differently. The accretion rate of polytropic fluids achieve a maximum at a negative value of $\Lambda$. Isothermal fluids accrete with efficiency that monotonically decreases with the increase of $\Lambda$.

The interesting feature of polytropic accretion flows is that the accretion stops completely when the \textit{absolute value} of $\Lambda$ is large enough. In all isothermal and polytropic accretion flows we observe freezing when \textit{positive} $\Lambda$ is large enough. This is yet another indication, in agreement with \cite{karkowski_malec_2013}, that the cyclic universe scenario of Penrose \cite{Penrose2004} cannot work.

The last Section contains a concise summary and addresses main issues of this paper.

Throughout this paper we adhere to the common relativistic notation. The determinant of the metric is denoted with $g = \mathrm{det} \, g_{\mu\nu}$. Greek indices run through $\mu = 0, 1, 2, 3$; Latin indices denote spatial dimensions $i = 1, 2, 3$. The signature of the metric is $(-,+,+,+)$, and we use geometric units with $G = c = 1$.

\section{Accretion in Schwarzschild--(anti-)de Sitter spacetimes}

\subsection{General equations}

We consider a static, spherically symmetric metric $g_{\mu \nu}$ with the line element of the form
\begin{equation}
\label{zxc}
ds^2 = g_{tt} dt^2 + 2 g_{tr} dt dr + g_{rr} dr^2 + R^2 \left( d\theta^2 + \sin^2 \theta d \phi^2 \right),
\end{equation}
where $g_{tt}$, $g_{tr}$, $g_{rr}$ and $R$ are functions of coordinate radius $r$ only. Angle variables satisfy $0\le \theta \le \pi $ and $0\le \phi < 2\pi$.

The motion of the fluid is described by standard conservation laws
\begin{equation}
\label{za}
 \nabla_\mu (\rho u^\mu) = \frac{1}{\sqrt{-g}} \partial_\mu \left( \sqrt{-g} \rho u^\mu \right) = 0,
\end{equation}
\begin{equation}
\label{zb}
\nabla_\mu \left( (e + p) u^\mu u^\nu + p g^{\mu \nu} \right) = 0.
\end{equation}
Here $u^\mu$ denotes the four-velocity of the fluid, $\rho$ is the baryonic density, $p$ is the pressure, and $e$ denotes the energy density. In the following, we will also use the specific enthalpy $h = (e + p)/\rho$.

We impose spherical symmetry onto hydrodynamic quantities. Thence $u^\theta = u^\phi = 0$, and $\rho , p, e$ and $h$  are functions of $r$ only.

The continuity equation (\ref{za}) can be integrated, yielding
\begin{equation}
\label{ad}
R^2 \sqrt{-g_{tt} g_{rr} + (g_{tr})^2 } \rho u^r = \mathrm{const}.
\end{equation}

We will further assume that the flow is smooth (i.e., no shock waves occur). Euler equations (\ref{zb}) can be then written as
\begin{equation}
\label{aa}
u^\mu \nabla_\mu \left( h u_\nu \right) + \partial_\nu h = u^\mu \partial_\mu(h u_\nu) - h \Gamma^\alpha_{\mu \nu} u^\mu u_\alpha + \partial_\nu h  = 0,
\end{equation}
where we have used the continuity equation (\ref{za}) and the thermodynamic relation $dh = dp / \rho$ (this assumes an isentropic flow --- cf.~\cite{landau}).  The zeroth component of the above equation (for $\nu = t$) reads
\[ u^r \partial_r \left( h u_t \right) - h \left( u^t u_t \Gamma^t_{tt} + u^t u_r \Gamma^r_{tt} + u^r u_t \Gamma^t_{rt} + u^r u_r \Gamma^r_{rt} \right) = 0. \]
All terms containing Christoffel symbols in the above expression cancel, and one is left with
\begin{equation}
\label{ab}
\partial_r (h u_t) = 0,
\end{equation}
or
\begin{equation}
\label{ac}
h u_t = \mathrm{const}.
\end{equation}
Exploiting the normalization condition for the four-velocity ($u_\mu u^\mu = -1$) one can show that
\[ u_t = - \sqrt{- g_{tt} + [ - g_{tt} g_{rr} + (g_{tr})^2 ]  (u^r)^2}. \]

For the barotropic equation of state $h = h(\rho)$ the solutions describing the flow can be found by solving the set of algebraic equations (\ref{ad}) and (\ref{ac}).

\subsection{Horizons and accretion in Schwarzschild--(anti-)de Sitter spacetimes}
 
In the following we search for solutions describing steady, spherically symmetric accretion flows in Schwarzschild--(anti-)de Sitter spacetimes. 

We choose the areal radius $R$ as the coordinate radius $r$. There is still a gauge freedom --- one can impose an arbitrary condition onto the trace $\mathrm{tr} K \equiv K_i^i$ of the extrinsic curvature $K^i_j$ of the Cauchy hypersurface $t = \mathrm{const}$ (see \cite{MTW} for the definition of extrinsic curvature) --- but in all gauges one has $- g_{tt} g_{rr} + (g_{tr})^2 = 1$. That plays a role in our forthcoming calculations; in all gauges
\[ u_t = - \sqrt{- g_{tt} + (u^r)^2}. \]

In standard coordinates all components of the extrinsic curvature vanish, and the line element is diagonal; it reads
\begin{equation}
ds^2 = - \left( 1 - \frac{2m}{r} - \frac{\Lambda}{3} r^2 \right) dt^2 + \frac{dr^2}{\left( 1 - \frac{2m}{r} - \frac{\Lambda}{3} r^2 \right)} + r^2 \left( d \theta^2 + \sin^2 \theta d \phi^2 \right).
\label{aj}
\end{equation}
Here $m$ denotes the Abbott--Deser mass of the central black hole \cite{Abbott-Deser}.

The horizons are located at zeros of $1 - 2m/r - (\Lambda/3)r^2$, or equivalently $-\frac{\Lambda}{3}r^3 + r - 2m$. For $0 < \Lambda < 1/(9 m^2)$ the cubic polynomial  $-\frac{\Lambda}{3}r^3 + r - 2m$ has 2 real positive zeros given by
\begin{eqnarray*}
r_a & = & \frac{2}{\sqrt{\Lambda}} \cos \left[ \frac{\pi}{3} + \frac{1}{3} \mathrm{arc \, cos} \left( 3 m \sqrt{\Lambda} \right) \right], \\
r_b & = & \frac{2}{\sqrt{\Lambda}} \cos \left[ \frac{\pi}{3} - \frac{1}{3} \mathrm{arc \, cos} \left( 3 m \sqrt{\Lambda} \right) \right].
\end{eqnarray*}
It is easy to show that
\[ 0 < 2m < r_a < 3m < \frac{1}{\sqrt{\Lambda}} < r_b < \frac{3}{\sqrt{\Lambda}}. \]
Here $r_a$ is the areal radius of the black-hole horizon, which will be further denoted by $r_h$, i.e., $r_h = r_a$; $r_b$ corresponds to the cosmological horizon. In this paper we will only deal with black holes satisfying $1 - 9 \Lambda m^2 > 0$. 

For $\Lambda < 0$ there is just one real root of  $1 - 2m/r - (\Lambda/3)r^2$ that corresponds to the black hole horizon. It is given by
\[ r_h = \frac{2}{\sqrt{|\Lambda|}} \sinh \left[ \frac{1}{3} \mathrm{ar \, sinh} \left( 3 m \sqrt{|\Lambda|} \right) \right]. \]

Equations (\ref{ad}) and (\ref{ac}) written in coordinates (\ref{aj}) have the form
\begin{equation}
\label{ae}
r^2 \rho u^r = \mathrm{const},
\end{equation}
\begin{equation}
\label{af}
h \sqrt{ 1 - \frac{2m}{r} - \frac{\Lambda}{3} r^2 + \left( u^r \right)^2 } = \mathrm{const}.
\end{equation}

Remarkably, it is possible to introduce new, suitably adapted Eddington--Finkelstein type coordinates that are regular at the black-hole horizon, and for which Eqs.~(\ref{ad}) and (\ref{ac}) have precisely the same form --- given by Eqs.~(\ref{ae}) and (\ref{af}). We define Eddington--Finkelstein time $t_\mathrm{EF}$ by
\[ dt = dt_\mathrm{EF} - \frac{\frac{2m}{r} + \frac{\Lambda}{3} r^2}{1 - \frac{2m}{r} - \frac{\Lambda}{3} r^2} dr. \]
In the new coordinates the metric can be written as
\[ ds^2 = - \left( 1 - \frac{2m}{r} - \frac{\Lambda}{3} r^2 \right) dt_\mathrm{EF}^2 + 2 \left( \frac{2m}{r} + \frac{\Lambda}{3} r^2 \right) dt_\mathrm{EF} dr +  \left( 1 + \frac{2m}{r} + \frac{\Lambda}{3}  \right)dr^2   + r^2 \left( d \theta^2 + \sin^2 \theta d \phi^2 \right). \]
The above coordinate system is singular for $\Lambda < 0$ and large values of $r$.

\subsection{Characteristics of the sonic point}

Equations (\ref{ae}) and (\ref{af}) admit transonic solutions --- the ones that are subsonic far from the center, and become supersonic in the vicinity of the black hole. We define the sonic point as a location in which $a^2 = (u^r/u_t)^2$, where $a$ is the local speed of sound .

By differentiating Eqs.~(\ref{ae}) and (\ref{af}) with respect to $r$ one can show that
\[ \left[ \left( \frac{u^r}{u_t} \right)^2 - a^2 \right] \partial_r \ln u^r = \frac{1}{r (u_t)^2} \left[ 2 a^2 (u_t)^2 - \frac{m}{r} + \frac{\Lambda}{3} r^2 \right] \]
for a barotropic equation of state of the form $h = h(\rho)$ (in this case $d h /h = a^2 d \rho / \rho$). Thus, if $a^2 = (u^r/u_t)^2$, and additionally $\partial_r \ln u^r$ is finite, we have
\[ 2 a^2 (u_t)^2 - \frac{m}{r} + \frac{\Lambda}{3} r^2 = 0, \]
and finally
\[ (u^r)^2 = \frac{m}{2r} - \frac{\Lambda}{6} r^2. \]
The assumption that $|\partial_r \ln u^r| < \infty$ is important. In the following we will describe a class of solutions that violate this condition, although $a^2 = (u^r/u_t)^2$.

In the rest of this paper, quantities referring to the sonic point will be denoted with an asterisk, i.e.,
\begin{equation}
\label{al}
(u_\ast^r)^2 = \frac{m}{2r_\ast} - \frac{\Lambda}{6} r^2_\ast = a_\ast^2 \left( 1 - \frac{3m}{2r_\ast} - \frac{\Lambda}{2} r_\ast^2 \right).
\end{equation}

\subsection{Boundary conditions}

We assume that the ball of fluid extends up to a radius $r_\infty$, where it has the baryonic density $\rho_\infty$ (or the energy density $e_\infty$) and the local speed of sound $a_\infty$. These data specify a transonic solution uniquely, with the exception of  possible bifurcations  --- see the discussion concerning ``homoclinic-type'' polytropic solutions). We demand also that the  boundary values satisfy $(u^r_\infty)^2 \ll 2 m /r_\infty \ll a_\infty^2$.

In the asymptotically flat fixed spacetime (e.g., in the Schwarzschild spacetime), one usually lets $r_\infty \to \infty$, so that $\rho_\infty$ and $a_\infty$ are asymptotic values. This defines a model that is mathematically elegant, but suffers from a physical contradiction --- the self-gravity of an infinite mass of fluid is neglected. A solution is to keep $r_\infty$ large, but finite. In this case it is also possible to take into account the self-gravity of the fluid, as it was done in \cite{karkowski_malec_2013}. In general, the assumption that the fluid region is finite is also necessary in Schwarzschild--(anti-)de Sitter spacetimes. Global solutions (with $r_\infty = \infty$) exist for $\Lambda < 0$ and isothermal equations of state.

\subsection{Mass accretion rates}

In the case of spherically symmetric, asymptotically flat systems with fluids one can define two sensible measures of mass. Thus it is not surprising that the mass accretion rate appears in two forms \cite{karkowski_malec_2013, kkmms}.

There is the quasilocal baryonic mass
\[ m_B(\Omega) = \int_\Omega d^3 x \sqrt{-g} u^t \rho; \]
here $\Omega$ is a 3-dimensional region of a $t = \mathrm{const}$ hypersurface. This definition follows directly from the continuity equation (\ref{za}); one has
\[ \partial_t m_B(\Omega) = - \int_{\partial \Omega} \sqrt{-g} \rho u^i dS_i, \]
where $dS_i$ denotes the integration element, normal to the boundary $\partial \Omega$. For the line element (\ref{aj}) the baryonic mass comprised within an anulus $(r_1,r_2)$ can be expressed as
\[ m_B(r_1,r_2) = 4 \pi \int_{r_1}^{r_2} dr r^2 u^t \rho. \]
It changes according to $\partial_t m_B(r_1,r_2) = \dot B(r_1) - \dot B(r_2)$. Here $\dot B = - 4 \pi r^2 u^r \rho$ is the baryonic mass accretion rate.

The other quasilocal measure of the mass contained in the anulus $(r_1,r_2)$ can be written, in coordinates (\ref{aj}), as
\begin{equation}
\label{wxxz}
m(r_1,r_2) = 4 \pi \int_{r_1}^{r_2} dr r^2 e.
\end{equation}
The corresponding mass accretion rate reads
\[ \dot m = - 4 \pi r^2 \sqrt{1 - \frac{2m}{r} - \frac{\Lambda}{3} r^2 + (u^r)^2} u^r (e + p). \]
Formula (\ref{wxxz}) is well known from the analysis of static systems in the polar gauge. It holds for accreting systems discussed in this paper as well. In fact, for self-gravitating spherically symmetric systems, this mass can be expressed as the surface-type Podurets--Misner--Sharp--Hawking \cite{Podurets} functional
\[ m(R) = \frac{R}{2}\left( 1 - \frac{R^2}{4} \theta (R) \theta^\prime (R)\right). \] 
The above expression can be also written as
\[ m(R) =   m - 4 \pi \int_R^{R_\infty} dr r^2 \left(  e - \frac{\mathrm{tr} K - K^r_r}{2k} n^i j_i \right) .\]
Here we assume the general metric of the form (\ref{zxc}). In this formulation $m$ is the conserved Abbott--Deser mass \cite{Abbott-Deser} in spacetimes that are asymptotically Schwarzschild--(anti-)de Sitter; $\theta = 2k + \mathrm{tr} K - K^r_r,  \theta^\prime = 2k - \mathrm{tr} K + K^r_r,$ denote the optical scalars of the surface $R = \mathrm{const}$; $j$ is the current density of the material field, while $n^\nu$ and $k$ are respectively a normal to a centered sphere given by $R = \mathrm{const}$ and its mean curvature, in the $t = \mathrm{const}$ hypersurface. This last formula has been obtained in \cite{MOM} for maximal hypersurfaces with $\mathrm{tr} K = 0$ of asymptotically flat geometries, but it is valid in any space-time slicing \cite{EM94, SH96}.
 
It has been found earlier, for $\Lambda = 0$, that both $\dot m$  and $\dot B$ are constant on a fixed time slice ($\partial_r \dot m = \partial_r \dot B = 0$ \cite{malec_1999, kkmms}). Since they do not depend on time in the steady flow approximation, they are proportional, $\partial_r \dot m = C \partial_r \dot B$, with a constant $C$ that can be found from boundary conditions.

Similarly one can obtain from Eqs.~(\ref{ae}) and (\ref{af}) that also for $\Lambda \neq 0$ neither $\dot B$ nor $\dot m$ depends on $r$. In the steady flow approximation we again have $\dot m=C\left( \Lambda \right) \dot B$, but now the constant depends on the cosmological constant. 

In the case of static systems the binding energy can be defined as the difference between the asymptotic and baryonic mass, $M_{bind} = m_B - m$ (\cite{Tooper}; see also \cite{km2004}). Then the sign of binding energy is correlated with the stability/instability of a spherical fluid ball; positive sign corresponds to stable systems, while negative binding energy indicates instability \cite{Tooper}. We conjecture that also in the steady flow approximation it is sensible to define \textit{the flow binding energy} by $M_{BF} \equiv \dot B - \dot m$. Thence locally $M_{BF} = \dot B (1 - C)$. Since $C$ changes with the change of $\Lambda$, that may induce a change in the sign of $M_{BF}$, possibly indicating  a change of its stability status.  

\section{Accretion of test polytropic fluids}

\subsection{Numerical solutions}

\begin{figure}
\setlength{\tabcolsep}{0.02\textwidth}
\begin{tabular}{@{}p{0.48\textwidth}p{0.48\textwidth}@{}}
\includegraphics[width=0.45\textwidth]{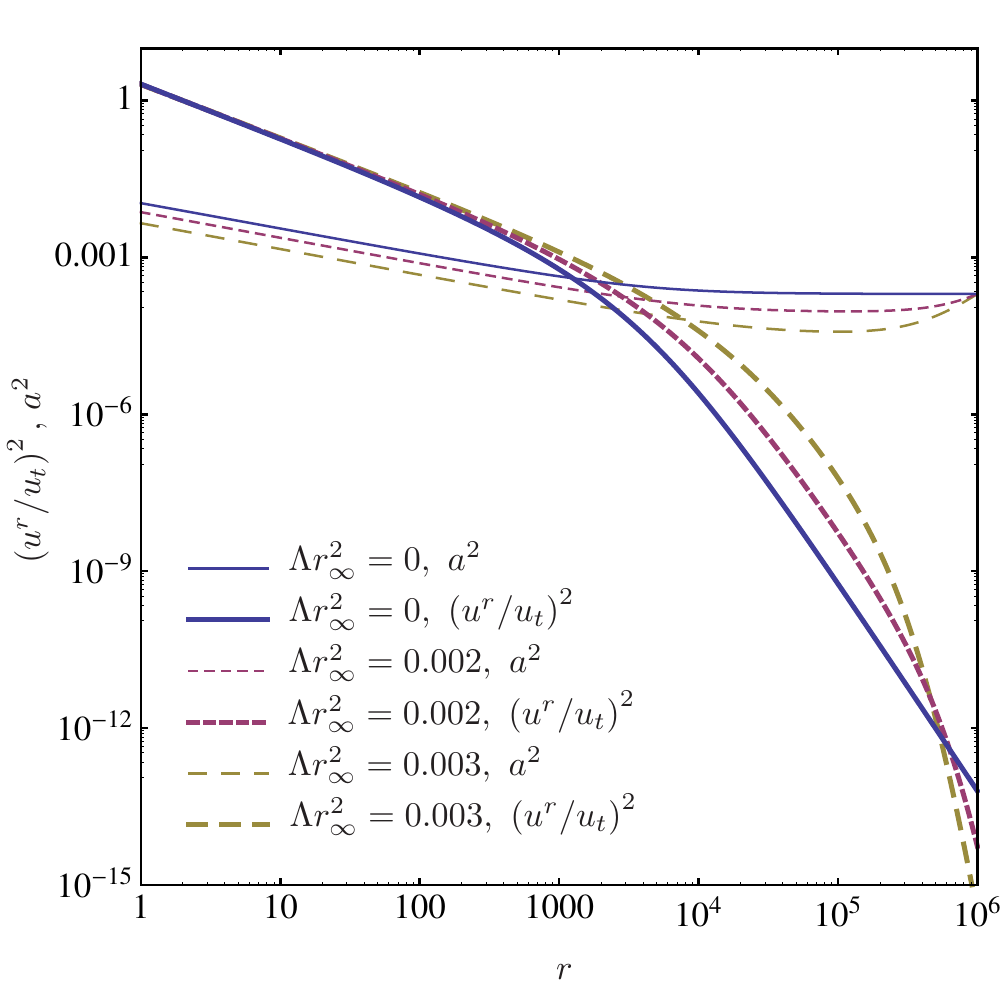}
&
\includegraphics[width=0.45\textwidth]{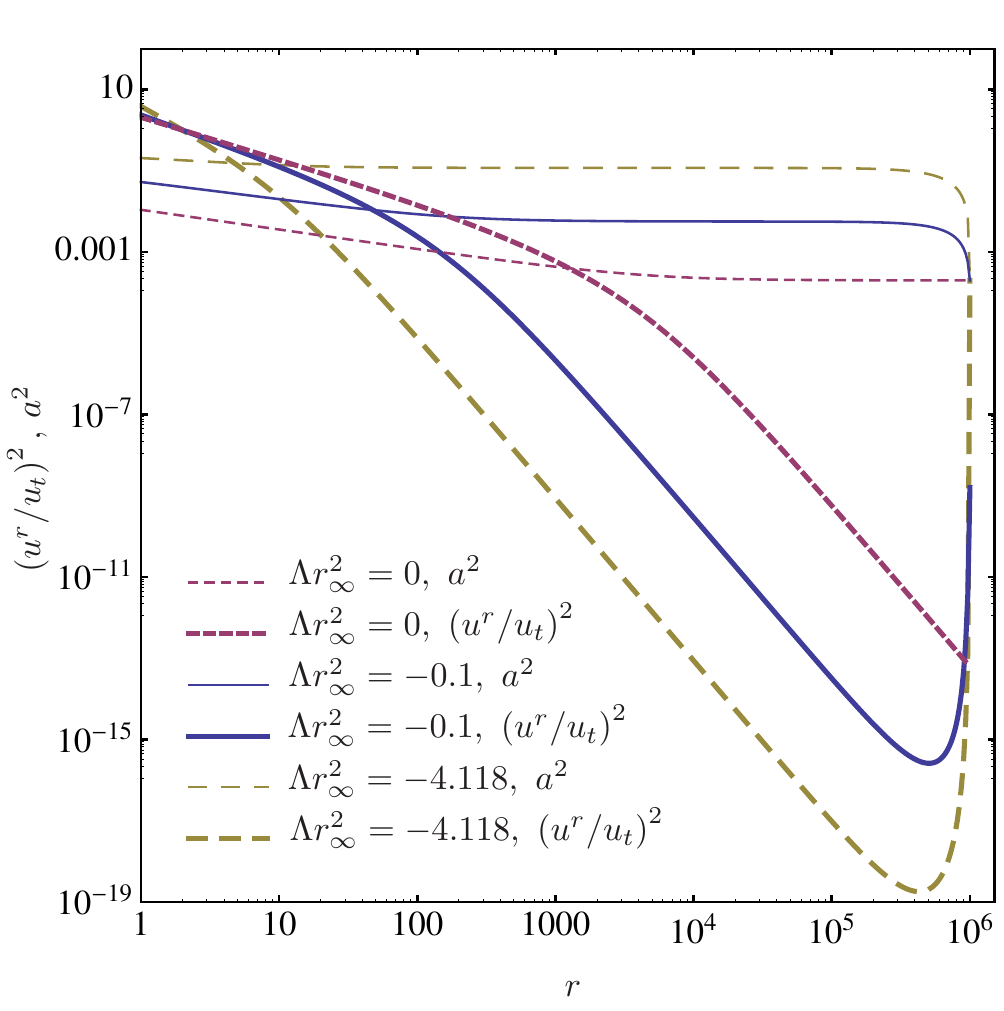}\\
\caption{\label{figa} Transonic solutions obtained for the polytropic equation of state with $\Gamma = 4/3$, $m = 1$, $r_\infty = 10^6$ and $a_\infty^2 = 2 \times 10^{-4}$. The plot shows graphs of both $a^2$ and $(u^r/u_t)^2$ for three different values of the cosmological constant $\Lambda r_\infty^2 = 0, 2 \times 10^{-3}$ and $3 \times 10^{-3}$.}
&
\caption{\label{figb} Same as in Fig.~\ref{figa}, but for $\Lambda r_\infty^2 = 0, -0.1$ and $-4.118$.}
\end{tabular}
\end{figure}

\begin{figure}
\includegraphics[width=0.45\textwidth]{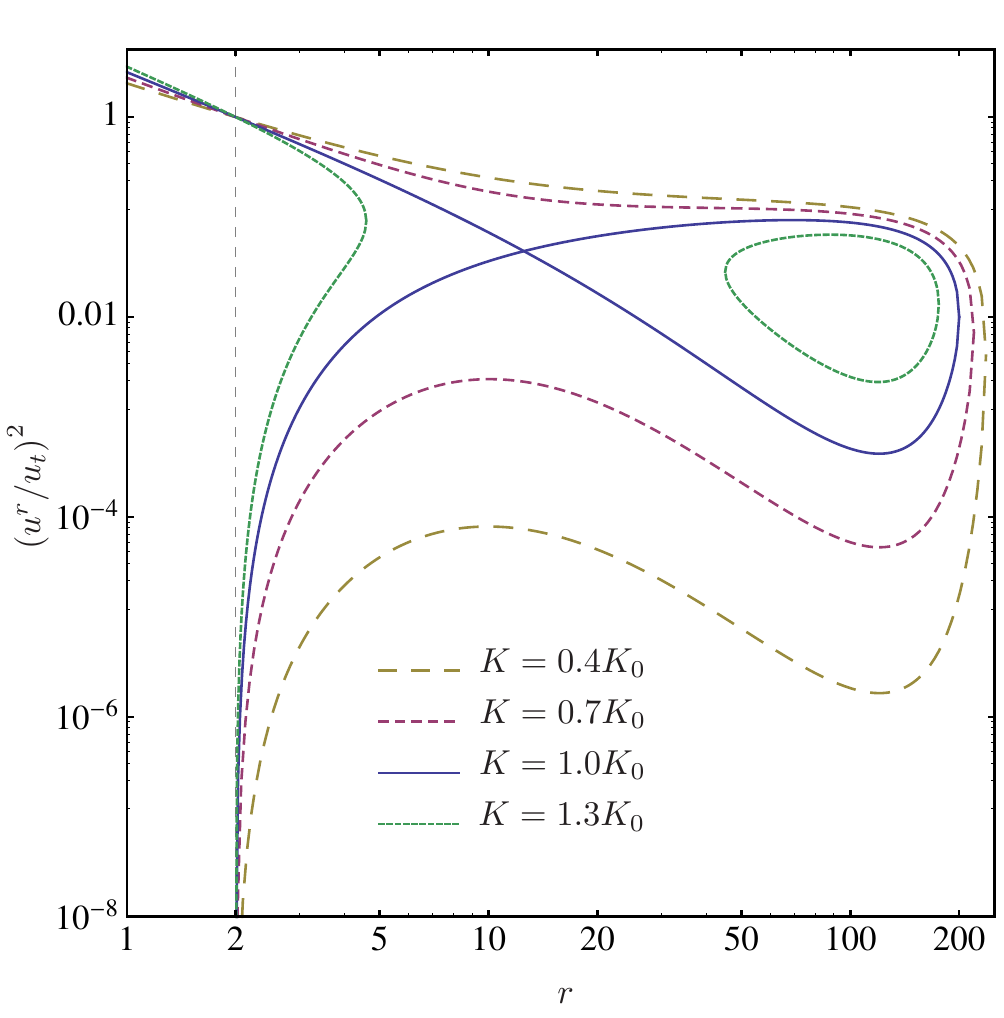}
\caption{\label{fig2} Solutions obtained for the polytropic equation of state with $\Gamma = 4/3$. Different curves correspond to solutions with different polytropic (entropy) constants $K$. The ``homoclinic'' solution (polytropic constant $K_0$) is characterized by the square of the speed of sound equal $a_\infty^2 = 1/100$ at $r_\infty = 200$. In this example $m = 1$, $\Lambda \approx - 0.3328$.}
\end{figure}

In this section we specialize to polytropic fluids obeying an equation of state of the form $p = K \rho^\Gamma$, where $K$ and $\Gamma$ are constant. For the polytropic equation of state
\[ h = \frac{\Gamma - 1}{\Gamma - 1 - a^2}, \]
and thus Eq.~(\ref{af}) can be written as
\[ (\Gamma - 1 - a^2) \sqrt{1 - \frac{2m}{r_\infty} - \frac{\Lambda}{3} r_\infty^2 + (u_\infty^r)^2} = (\Gamma - 1 - a_\infty^2) \sqrt{1 - \frac{2m}{r} - \frac{\Lambda}{3} r^2 + (u^r)^2}. \]
Similarly
\begin{equation}
\label{ag}
\rho = \rho_\infty \left( \frac{a^2}{a_\infty^2} \frac{\Gamma - 1 - a_\infty^2}{\Gamma - 1 - a^2} \right)^\frac{1}{\Gamma - 1},
\end{equation}
and
\begin{equation}
\label{ah}
u^r = \frac{r_\infty^2}{r^2} \left( \frac{a_\infty^2}{a^2} \frac{\Gamma - 1 - a^2}{\Gamma - 1 - a_\infty^2} \right)^\frac{1}{\Gamma - 1} u^r_\infty.
\end{equation}

Parameters characterizing a transonic solution are obtained by solving, with respect to $r_\ast$, equation
\begin{widetext}
\begin{equation}
\label{ba}
(\Gamma - 1 - a_\ast^2) \sqrt{1 - \frac{2m}{r_\infty} - \frac{\Lambda}{3} r_\infty^2 + (u_\ast^r)^2 \frac{r_\ast^4}{r_\infty^4} \left( \frac{a_\ast^2}{a_\infty^2} \frac{\Gamma - 1 - a_\infty^2}{\Gamma - 1 - a_\ast^2} \right)^\frac{2}{\Gamma - 1} } = (\Gamma - 1 - a_\infty^2) \sqrt{1 - \frac{3m}{2r_\ast} - \frac{\Lambda}{2} r_\ast^2},
\end{equation}
\end{widetext}
where we substitute
\[ (u_\ast^r)^2 = \frac{m}{2r_\ast} - \frac{\Lambda}{6} r^2_\ast \]
and
\[ a_\ast^2 = \frac{ \frac{m}{2r_\ast} - \frac{\Lambda}{6} r^2_\ast }{ 1 - \frac{3m}{2r_\ast} - \frac{\Lambda}{2} r_\ast^2 }. \]
In general, this procedure yields many roots, from which we select those that are real and positive. In the next step one computes all values of $a_\ast^2$ corresponding to obtained values of $r_\ast$. A general restriction for $a_\ast^2$ is $0 < a_\ast^2 < \Gamma - 1$. In the last step one has to ensure that the solution satisfy
\[ \frac{\left( u_\infty^r \right)^2}{1 - \frac{2m}{r_\infty} - \frac{\Lambda}{3} r_\infty^2 + \left( u_\infty^r \right)^2 } < a_\infty^2, \]
i.e., that the solution is subsonic at the outer boundary. The asymptotic velocity $u_\infty$ can be obtained from
\[ u_\infty^r = \frac{r_\ast^2}{r_\infty^2} \left( \frac{a_\ast^2}{a_\infty^2} \frac{\Gamma - 1 - a_\infty^2}{\Gamma - 1 - a_\ast^2} \right)^\frac{1}{\Gamma - 1} u_\ast^r. \]
Note that the only parameters that enter the above equations are $m$, $\Lambda$, $r_\infty$ and $a_\infty$, i.e., the properties of the sonic point do not depend on $\rho_\infty$.

We obtain the solution by solving the equation
\[
(\Gamma - 1 - a^2) \sqrt{1 - \frac{2m}{r_\infty} - \frac{\Lambda}{3} r_\infty^2 + (u_\infty^r)^2} =
 (\Gamma - 1 - a_\infty^2) \sqrt{1 - \frac{2m}{r} - \frac{\Lambda}{3} r^2 + \frac{r_\ast^4}{r^4} \left( \frac{a_\ast^2}{a^2} \frac{\Gamma - 1 - a^2}{\Gamma - 1 -a_\ast^2} \right)^\frac{2}{\Gamma - 1} \left( u_\ast^r \right)^2 }
\]
with respect to $a^2$ in the interesting range of $r$. Finally, $u^r$ is computed as
\[ u^r = \frac{r_\ast^2}{r^2} \left( \frac{a_\ast^2}{a^2} \frac{\Gamma - 1 - a^2}{\Gamma - 1 -a_\ast^2} \right)^\frac{1}{\Gamma - 1} u_\ast^r.  \]

Sample solutions corresponding to the parameters that were specified in \cite{karkowski_malec_2013}, i.e., $m =1$, $\Gamma = 4/3$, $r_\infty = 10^6$ and $a_\infty^2 = 2 \times 10^{-4}$, are plotted in Figs.~\ref{figa} and \ref{figb} for positive and negative $\Lambda$, respectively. Both plots show graphs of $(u^r/u_t)^2$ and $a^2$, intersecting at sonic points. Figure \ref{figa} depicts solutions obtained for $\Lambda r_\infty^2 = 0, 2 \times 10^{-3}$ and $3 \times 10^{-3}$. Solutions shown in Fig.~\ref{figb} correspond to $\Lambda r_\infty^2 = 0, -0.1$ and $-4.118$.

It is important to stress that in both sectors (of positive and negative $\Lambda$) there are limits on the maximum absolute value of the cosmological constant for which stationary solutions exist. This would be a natural behavior to expect in the case of Schwarzschild--de Sitter spacetime, where the location of the cosmological horizon is shifted towards lower and lower values with an increase of $\Lambda$, and it coincides with the outer boundary of the accreting cloud for $\Lambda r_\infty^2 = 3 \left( 1 - 2 m/r_\infty \right)$. In fact, for polytropic fluids, stationary solutions cease to exist for much lower values of $\Lambda$. In the example specified above, no stationary solution exist for $\Lambda r_\infty \gtrsim 3.69 \times 10^{-3} $.

It is quite surprising that there is also a limit on the maximal allowed $|\Lambda|$ in the Schwarzschild--anti-de Sitter case. An interesting mechanism is responsible for this fact, and it can be observed already in Fig.~\ref{figb}. Here, for $\Lambda r_\infty^2 \approx -4.118$ the value of $(u^r/u_t)^2$ at $r = r_\infty$ reaches $2 \times 10^{-4}$, that is precisely the assumed value of $a_\infty^2$, and the graphs of $(u^r/u_t)^2$ and $a^2$ form a closed loop. No stationary transonic solutions can be found for larger $|\Lambda|$. The limiting solution is a part of what we call a ``homoclinic-type'' solution. By ``homoclinic-type'' solutions we mean solutions that form closed loops on the graph of $(u^r/u_t)^2$ versus $r$, joining a single critical sonic point. In this respect they are similar to homoclinic orbits in the theory of dynamical systems.

In the standard picture of Bondi-type accretion two stationary solutions intersect at a sonic point. One of them is subsonic outside the sonic point and becomes supersonic for $r < r_\ast$ (such solutions are plotted in Figs.~\ref{figa} and \ref{figb}). The other solution is supersonic for $r > r_\ast$ and subsonic for $r < r_\ast$, and it is usually interpreted as describing ``wind'' instead of accretion flow, although the equations do not specify the sign of the radial velocity. In the case of the limiting, ``homoclinic-type'' solution described above, this additional branch satisfies $(u^r/u_t)^2 = a^2$ at $r = r_\infty$, so the two branches of the solution join again at $r_\infty$. Note that in this case $|\partial_r u^r| = \infty$ at $r_\infty$. This situation is depicted in Fig.~\ref{fig2}. For the sake of clarity, the solutions in this plot were computed assuming exaggerated values of parameters. Unlike those plotted in Fig.~\ref{figb}, these solutions do not satisfy our assumptions $(u_\infty^r)^2 \ll 2m/r_\infty \ll a_\infty^2$. The ``homoclinic'' (or limiting) solution was obtained for $m = 1$, $\Gamma = 4/3$, $r_\infty = 200$, $a_\infty^2 = 10^{-2}$ and $\Lambda r_\infty^2 \approx - 0.3329$. 

The structure of solutions for $\Lambda < 0$ can be better understood, when inspecting the whole family of solutions characterized by the same constants that appear on the right-hand sides of Eqs.~(\ref{ae}) and (\ref{af}), i.e., solutions satisfying the same integral conservation laws. They can differ by the specific entropy, proportional to the polytropic constant $K$, and they correspond to different boundary values $a_\infty$ and $\rho_\infty$. Such solutions can be joined by stationary shock waves, provided that appropriate Rankine--Hugoniot conditions are satisfied. A sample of these solutions is plotted in Fig.~\ref{fig2}. They are computed assuming polytropic constants $K = 0.4 K_0$, $0.7 K_0$ and $1.3 K_0$, where $K_0$ denotes the polytropic constant corresponding to the ``homoclinic'' solution (this is the value that can be computed from the assumed boundary conditions).

\subsection{Mass accretion rates}

Given the numerical solutions described above, one can easily find the corresponding accretion rates. The baryonic mass accretion rate can be computed as
\begin{equation}
\dot B = 4 \pi |u^r_\infty | r_\infty^2 \rho_\infty = 4 \pi \sqrt{\frac{m}{2 r_\ast} - \frac{\Lambda}{6}r_\ast^2} r_\ast^2 \left( \frac{a_\ast^2}{a_\infty^2} \frac{\Gamma - 1 - a_\infty^2}{\Gamma - 1 - a_\ast^2} \right)^\frac{1}{\Gamma - 1}  \rho_\infty.
\label{bb}
\end{equation}
One can also obtain a compact expression for $\dot m$. A straightforward calculation gives
\begin{equation}
\label{nnbb}
\dot m = C \dot B, \quad \mathrm{where} \quad C =  \frac{\Gamma - 1}{\Gamma - 1 - a_\ast^2} \frac{\sqrt{1 - \Lambda r_\ast^2}}{\sqrt{1 + 3 a^2_\ast}}.
\end{equation}
We postpone the discussion of the resulting dependence of $\dot B$ and $\dot m$ on $\Lambda$ to Section V.

The binding energy flow reads $M_{BF} = \dot B - \dot m = \dot B \left(  1 - C \right)$. It is instructive to investigate the dependence of $C$ on $a^2_\ast $. One finds (remember that $r_\ast $ depends on $a_\ast $, through the sonic point relations)
\[ \frac{\partial C}{\partial a^2_\ast} = \frac{C}{1 + 3 a_\ast^2} \left[  \frac{5 - 3 \Gamma + 9 a_\ast^2 }{2 \left( \Gamma - 1 - a_\ast^2 \right)} + \frac{\Lambda r^2_\ast }{a^2_\ast \left(  1  - \frac{3}{2} \Lambda r^2_\ast \right) + \frac{1}{2} \Lambda r^2_\ast } \right] .\]
Thus, taking into account that $\Gamma \le 5/3$ and $|\Lambda |r^2_\ast \approx 0$, we infer that $C$ increases with $a^2_\ast$. Now $C(a^2_\ast \approx 0)$ is bigger than 1 if $\Lambda < 0$; thence $M_{BF} < 0$. In the opposite case, when $\Lambda > 0$ and $a^2_\ast \approx 0$, $M_{BF}$ is positive, but with the increase of $\Lambda$ it may become negative. If one can pursue the analogy with static polytropic fluids \cite{Tooper}, then $M_{BF} < 0$ would signal instability of the accreting system. This stability issue will be investigated elsewhere.

\section{Accretion of test isothermal fluids}

\begin{figure}
\setlength{\tabcolsep}{0.02\textwidth}
\begin{tabular}{@{}p{0.48\textwidth}p{0.48\textwidth}@{}}
\includegraphics[width=0.45\textwidth]{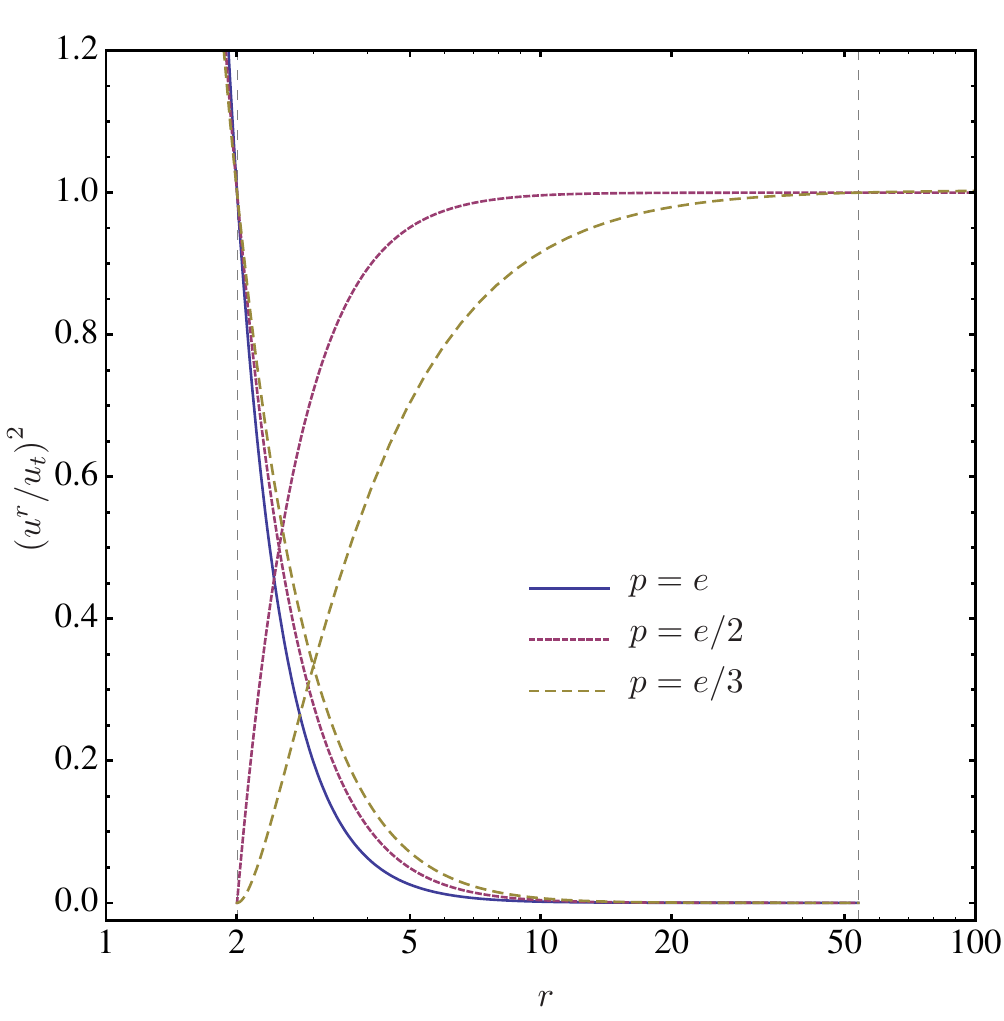}
&
\includegraphics[width=0.45\textwidth]{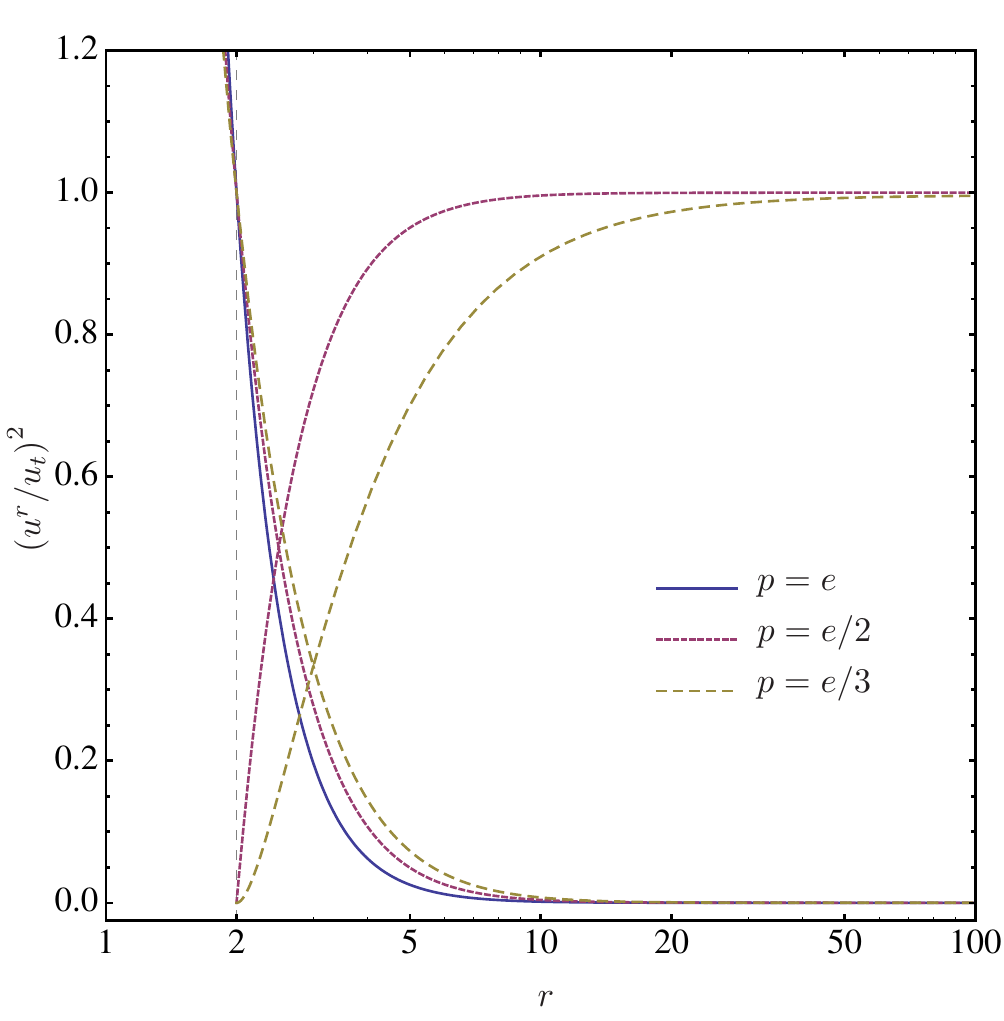}\\
\caption{\label{fig3} Solutions obtained for equations of state $p = k e$, with $k = 1/3, 1/2, 1$, $\Lambda = 1/1000$ and $m = 1$. Dotted vertical lines denote the locations of the horizons.}
&
\caption{\label{fig4} Solutions obtained for equation of state $p = k e$, with $k = 1/3, 1/2, 1$, $\Lambda = -1/1000$ and $m = 1$. The dotted vertical line denotes the location of the horizon.}
\end{tabular}
\end{figure}

For the isothermal equation of state $p = ke$ the ``baryonic'' density can be computed as
\begin{equation}
\label{ccab}
\rho = \rho_\infty \exp \int_{e_\infty}^e \frac{de^\prime}{e^\prime + p(e^\prime)} = \rho_\infty  \left( \frac{e}{e_\infty} \right)^\frac{1}{1+k}.
\end{equation}
The specific enthalpy is
\begin{equation}
\label{ccg}
h = \frac{e + p}{\rho} = \frac{(1+k) e_\infty}{\rho_\infty^{1+k}} \rho^k.
\end{equation}
Here $e_\infty$ denotes the value of the energy density $e$ at the outer boundary of the cloud. Note that the above calculation is formal, and it should be properly interpreted. Here $\rho$ is understood as an integrability factor for which the current $\rho u^\mu$ is conserved in the smooth region of the flow. For the fluid with zero baryonic density (the Harrison et al. limit \cite{Harrison}) ---  e.g. photon gas with $p = e/3$ --- a non-trivial quantity of this type is the entropy, and the proper physical interpretation of $\rho u^\mu$ is that of the entropy current. Also note that it is not conserved across possible shock waves (see, e.g., \cite{mach_pietka}).

Eqs.~(\ref{ae}) and (\ref{af}) can be now written as
\begin{equation}
\label{gb}
r^2 \rho u^r = \mathrm{const},
\end{equation}
\begin{equation}
\label{gc}
\rho^k \sqrt{ 1 - \frac{2m}{r} - \frac{\Lambda}{3} r^2 + \left( u^r \right)^2 } = \mathrm{const}.
\end{equation}
For $k = 1/3, 1/2, 1$ the above set of equations can be reduced to a polynomial equation of the order $\le 3$, and one can obtain explicit solutions. The case with $k = 1/3$ is known as the  radiation   fluid --- it can be viewed as the limit of a perfect gas equation of state with $\Gamma = 4/3$ in which the contribution of the baryonic density to the energy density is negligible. The case with $k = 1$ yields a particularly elegant and simple solution. Accretion of $k=1$ fluid  onto a moving  Schwarzschild or Kerr black hole was solved   in \cite{petrich}.

Accretion of the fluid with the so-called ``phantom'' equations of state of the form $p = k\rho$, $k < 0$ in Schwarzschild--(anti-)de Sitter spacetimes was also considered in \cite{amani_farahani, sharif_abbas}, but with no analysis of explicit solutions.

\subsection{Sonic points}

Transonic solutions can be also constructed for equations of state of the form $p = k e$. Note that $a^2 = k = \mathrm{const}$, and Eq.~(\ref{al}) yields
\begin{equation}
\label{ga}
(u_\ast^r)^2 = \frac{m}{2r_\ast} - \frac{\Lambda}{6} r^2_\ast = k \left( 1 - \frac{3m}{2r_\ast} - \frac{\Lambda}{2} r_\ast^2 \right).
\end{equation}
The above equation can be solved for $r_\ast$ assuming any value of $k$, but we will only need solutions for $k = 1/3, 1/2, 1$.

Setting $k = 1/3$ yields a unique solution $r_\ast = 3m$, $(u_\ast^r)^2 = (1 - 9 \Lambda m^2)/6$. Note that the location of the sonic point does not depend on the cosmological constant!

For $k = 1$, Eq.~(\ref{ga}) can be reduced to $1 - 2m/r_\ast - (\Lambda/3) r_\ast^2 = 0$, i.e., the sonic point (or sonic points) is (are) located at the horizon(s).

For $k = 1/2$, the location of the sonic point can be obtained by solving the equation
\begin{equation}
\label{ha}
\frac{\Lambda}{6} r_\ast^3 - r_\ast + \frac{5}{2} m = 0.
\end{equation}
For $\Lambda < 0$, the discriminant of the above equation, that is
\[ \frac{1}{\Lambda^2} \left( \left( \frac{15m}{2} \right)^2 - \frac{2^3}{\Lambda} \right), \]
is positive, and the only real root is given by
\begin{widetext}
\[ r_\ast  =  2 \sqrt{\frac{2}{|\Lambda|}} \sinh \left( \frac{1}{3} \mathrm{ar\, sinh} \left( \frac{15 m \sqrt{|\Lambda|}}{4 \sqrt{2}} \right) \right). \]
\end{widetext}
For $\Lambda > 0$, the discriminant is negative (remember that $1 - 9 \Lambda m^2 > 0$), and there are 3 real solutions. The position of the sonic point is given by
\[ r_{\ast 1} = 2 \sqrt{\frac{2}{\Lambda}} \cos \left( \frac{\pi}{3} + \frac{1}{3} \mathrm{arc \, cos} \left( \frac{15 m \sqrt{\Lambda}}{4 \sqrt{2}} \right) \right). \]
Another positive root of Eq.~(\ref{ha}), i.e.
\[ r_{\ast 2} = 2 \sqrt{\frac{2}{\Lambda}} \cos \left( \frac{\pi}{3} - \frac{1}{3} \mathrm{arc \, cos} \left( \frac{15 m \sqrt{\Lambda}}{4 \sqrt{2}} \right) \right), \]
is larger than the areal radius of the cosmological horizon. Note that $r_a \le r_{\ast 1} \le r_b < r_{\ast 2}$, where $r_a$ and $r_b$ are the locations of the horizons.

\subsection{Mass accretion rates}

For transonic flows the  mass accretion rate $\dot m$ can be expressed by a simple formula depending on $r_\ast$ and $k$. A straightforward calculation, making use of Eqs.~(\ref{ccg}), (\ref{gb}), (\ref{gc}) and (\ref{ga}), yields
\begin{equation}
\label{nncg}
\dot m =   \frac{4 \pi (1+k)}{\sqrt{k} r_\ast^{2k}} \left( \frac{m}{2 r_\ast} - \frac{\Lambda}{6} r_\ast^2 \right)^\frac{1-k}{2} \left( r_\infty^2 |u_\infty^r| \right)^{1+k} e_\infty.
\end{equation}

Note that, according to earlier arguments, $\dot B = 4 \pi \rho_\infty r_\infty^2 |u_\infty^r|$ should be rather interpreted as the entropy accretion rate. In the following, we will discuss the behavior of $\dot B$ mainly to maintain contact with previous works.

\subsection{Isothermal fluids: explicit solutions}

\subsubsection{Radiation fluids: $k = 1/3$}

For $k = 1/3$, Eqs.~(\ref{gb}) and (\ref{gc}) yield
\[ \frac{\left( 1 - \frac{2m}{r} - \frac{\Lambda}{3} r^2 + (u^r)^2 \right)^3}{r^4 (u^r)^2}  = \mathrm{const} = \frac{\left( 1 - \frac{2m}{r_\ast} - \frac{\Lambda}{3} r_\ast^2 + (u_\ast^r)^2 \right)^3}{r^4 (u_\ast^r)^2} = \frac{3}{4} \frac{(1 - 9 \Lambda m^2)^2}{(3m)^4}, \]
where we have used the obtained parameters of the sonic point. In the following we solve the above equation for $(u^r)^2$. Introducing $Y = 1 - 2m/r - (\Lambda/3)r^2 + (u^r)^2$, one obtains the equation $Y^3 - bY + ab = 0$, where
\[ a = 1 - \frac{2m}{r} - \frac{\Lambda}{3}r^2 \;\;\; \mathrm{and} \;\;\; b = \frac{3}{4} \frac{(1 - 9\Lambda m^2)^2}{(3m)^4} r^4. \]
One can show that the discriminant of the above cubic equation is non-positive, provided that $1 - 9 \Lambda m^2 > 0$. It vanishes for $r = 3m$, i.e., at the sonic point. Thus, except for the sonic point, there are 3 real solutions, but only 2 of them yield positive $(u^r)^2$. Defining
\begin{eqnarray*}
X_1 & = & - 1 + \frac{2m}{r} + \frac{\Lambda}{3}r^2 +  \frac{(1 - 9\Lambda m^2) r^2}{(3 m)^2} \cos \left\{ \frac{\pi}{3} - \frac{1}{3} \mathrm{arc \, cos} \left[ \frac{3^3 m^2 \left( 1 - \frac{2m}{r} - \frac{\Lambda}{3} r^2 \right)}{(1 - 9 \Lambda m^2) r^2} \right] \right\}, \\
X_2 & = & - 1 + \frac{2m}{r} + \frac{\Lambda}{3}r^2 + \frac{(1 - 9\Lambda m^2) r^2}{(3 m)^2} \cos \left\{ \frac{\pi}{3} + \frac{1}{3} \mathrm{arc \, cos} \left[ \frac{3^3 m^2 \left( 1 - \frac{2m}{r} - \frac{\Lambda}{3} r^2 \right)}{(1 - 9 \Lambda m^2) r^2} \right] \right\}, \\
\end{eqnarray*}
one can write the expressions for the two branches of the solution as
\begin{equation}
\label{zz1}
(u^r)^2 = \left\{\begin{array}{ll} X_2, & r \ge 3m, \\ X_1, & r < 3m, \end{array} \right.
\end{equation}
for the branch that is subsonic outside the sonic point, and
\begin{equation}
\label{zz2}
(u^r)^2 = \left\{\begin{array}{ll} X_1, & r \ge 3m, \\ X_2, & r < 3m, \end{array} \right.
\end{equation}
for the branch that is supersonic outside the sonic point.

For $\Lambda > 0$, both branches are well defined in the region enclosed between the horizon of the black hole (located at $r_a$) and the cosmological horizon (located at $r_b$). The solution given by Eq.~(\ref{zz1}) yields $u^r = 0$ at $r = r_b$, so that the density  $\rho$ becomes divergent at the cosmological horizon. Similarly, for the branch given by Eq.~(\ref{zz2}), one has $u^r = 0$ at $r = r_a$, and $\rho$ diverges at the horizon of the black hole.   On the other hand, the solution given by Eq.~(\ref{zz1}) can be continued beneath the horizon of the black hole, while the branch described by Eq.~(\ref{zz2}) extends beyond the cosmological horizon. For $0 < \Lambda m^2 < 1/18$ this solution can be continued arbitrarily far outside the cosmological horizon, but there is a limit on this continuation for $1/18 < \Lambda m^2 < 1/9$. In the latter case, there exist a finite radius $r > r_b$ for which the argument of the $\mathrm{arc \, cos}$ function appearing in Eq.~(\ref{zz2}) reaches $-1$.

 These facts can be interpreted in the following intuitively transparent way. Solution (\ref{zz1}) describes the accretion process; the accreting gas cannot extend beyond the cosmological horizon, while it can penetrate the interior of a black hole. Conversely, the solution given by Eq.~(\ref{zz2}) must describe the wind. Therefore it cannot start from the horizon of a black hole (our fluids satisfy the dominant energy conditions) but instead it can extend beyond the cosmological horizons. Obviously, one must choose the right  sign of $u^r$ --- positive for the wind and or negative for the accretion.

Both accretion rates $\dot B$ and $\dot m$ can be easily computed, yielding
\[ \dot B =4 \pi \rho_\infty r_\infty^2 |u_\infty^r| \]
and
\[ \dot m = \frac{16 \pi}{3 \sqrt{3} \sqrt[3]{2 m^2}} \left(1 - 9 \Lambda m^2 \right)^\frac{1}{3} \left( r_\infty^2 |u_\infty^r| \right)^\frac{4}{3} e_\infty, \]
where
 
\begin{equation}
\label{bcz}
r_\infty^2 |u_\infty^r| = r_\infty^2 \left\{ - 1 + \frac{2m}{r_\infty} + \frac{\Lambda}{3}r_\infty^2 + \frac{(1 - 9\Lambda m^2) r_\infty^2}{(3 m)^2} \cos \left[ \frac{\pi}{3} + \frac{1}{3} \mathrm{arc \, cos} \left( \frac{3^3 m^2 \left( 1 - \frac{2m}{r_\infty} - \frac{\Lambda}{3} r_\infty^2 \right)}{(1 - 9 \Lambda m^2) r_\infty^2} \right) \right] \right\}^\frac{1}{2}
\end{equation}
for a solution that is subsonic outside the sonic point. Note that if the equation of state with $k = 1/3$ is interpreted as describing the fluid with vanishing baryonic density, the expression for $\dot B$ gives the entropy accretion rate.

Both expressions can be simplified in the case where $r_\infty \gg m$. Equation (\ref{bcz}) can be expanded in powers of $m/r_\infty$ around $m/r_\infty = 0$, keeping the product $\Lambda r_\infty^2$ fixed. The zero-order term in this expansion is
\[ r_\infty^2 |u_\infty^r| = 6 \sqrt{3} \pi m^2 \left( 1 - \Lambda r_\infty^2/3 \right)^{3/2}, \]
and the accretion rates can be approximated as
\[ \dot B \approx 24 \sqrt{3} (\pi m)^2 \rho_\infty \left( 1- \Lambda  r_\infty^2/3 \right)^{3/2} \]
and
\[ \dot m \approx 32 \sqrt{3} (\pi m)^{2} \left( 1 - \Lambda r_\infty^2 / 3 \right)^2 \left(1 - 9 \Lambda m^2 \right)^{1/3} e_\infty. \]

\subsubsection{Solution for $k=1/2$}

In this case,  Eqs.~(\ref{gb}) and (\ref{gc}) yield --- expressing the square $(u^r)^2_\ast $ and $\Lambda r^2_\ast $ in terms of $m/r_\ast $    ---
 \begin{equation}
\label{ia}
\frac{1 - \frac{2m}{r} - \frac{\Lambda}{3} r^2 + (u^r)^2}{r^2 u^r} = \mathrm{const} = \frac{1 - \frac{2m}{r_\ast} - \frac{\Lambda}{3} r_\ast^2 + (u_\ast^r)^2}{r_\ast^2 u_\ast^r} = - \frac{2}{r_\ast^2} \sqrt{\frac{m}{2 r_\ast} - \frac{\Lambda}{6} r_\ast^2} = - \frac{2}{r_\ast^2} \sqrt{\frac{3m}{r_\ast} - 1},
\end{equation}
where we assumed that $u^r < 0$. Define
\begin{eqnarray*}
Y_1 & = &  B r^2 + \sqrt{B^2 r^4 - 1 + \frac{2m}{r} + \frac{\Lambda}{3} r^2}, \\
Y_2 & = &  B r^2 - \sqrt{B^2 r^4 - 1 + \frac{2m}{r} + \frac{\Lambda}{3} r^2},
\end{eqnarray*}
where
\[ B = - \frac{1}{r_\ast^2} \sqrt{\frac{3m}{r_\ast} - 1}, \]
and
\[ r_\ast = \left\{ \begin{array}{ll}
2 \sqrt{\frac{2}{|\Lambda|}} \sinh \left( \frac{1}{3} \mathrm{ar\, sinh} \left( \frac{15 m \sqrt{|\Lambda|}}{4 \sqrt{2}} \right) \right), & \Lambda < 0, \\
5m/2, & \Lambda = 0,\\
2 \sqrt{\frac{2}{\Lambda}} \cos \left( \frac{\pi}{3} + \frac{1}{3} \mathrm{arc \, cos} \left( \frac{15 m \sqrt{\Lambda}}{4 \sqrt{2}} \right) \right), & 0 < \Lambda < \frac{1}{9m^2}.
\end{array} \right. \]

The solution of Eq.~(\ref{ia}) can be written as
\begin{equation}
\label{zbb}
u^r = \left\{ \begin{array}{ll} Y_1, & r \ge r_\ast, \\ Y_2, & r < r_\ast, \end{array} \right.
\end{equation}
representing the branch that is subsonic outside the sonic point, and
\begin{equation}
\label{zaa}
u^r = \left\{ \begin{array}{ll} Y_2, & r \ge r_\ast, \\ Y_1, & r < r_\ast, \end{array} \right.
\end{equation}
for the branch that is supersonic outside the sonic point. Note that in both cases, taking $u^r$ with the reversed sign yields the solution representing ``wind'' instead of accretion.

The above solutions have a similar structure to those obtained for $k = 1/3$. The density corresponding to the velocity given by Eq.~(\ref{zaa}) is divergent at the horizon of the black hole;
clearly, it must represent the wind solution, according to the preceding discussion concerning the $k=1/3$ isothermal fluid. Similarly, solution (\ref{zbb}) has a divergent density at the cosmological horizon (for $\Lambda > 0$); this must describe the accretion process. Obviously, the sign of $u^r$ must be positive or negative, respectively.

The accretion rates corresponding to  solutions that are subsonic outside the sonic point read
\[ \dot B = 4 \pi \rho_\infty r_\infty^2 |u_\infty^r| \]
and
\[ \dot m = \frac{6 \sqrt{2} \pi}{r_\ast} \left( \frac{m}{2r_\ast} - \frac{\Lambda}{6} r_\ast^2 \right)^\frac{1}{4} \left( r_\infty^2 |u_\infty^r| \right)^\frac{3}{2} e_\infty, \]
where
\[ r_\infty^2 |u_\infty^r| = - r_\infty^2 \left( B r_\infty^2 + \sqrt{B^2 r_\infty^4 - 1 + \frac{2m}{r_\infty} + \frac{\Lambda}{3}r_\infty^2 } \right). \]
These expressions  can be put in  simpler (approximated) forms, when we employ the boundary condition $r_\infty \gg m$. In this case
\[ \dot B \approx 4 \pi \rho_\infty r_\ast^2 \frac{\left(1 - \frac{\Lambda}{3} r^2_\infty \right)}{\sqrt{\frac{3m}{r_\ast} -1} + \sqrt{ \frac{3m}{r_\ast} - 1  + \frac{\Lambda r_\ast^4}{3r_\infty^2}}} \]
and
\[ \dot m \approx 6 \sqrt{2} \pi \left( \frac{m}{2r_\ast} - \frac{\Lambda}{6} r_\ast^2 \right)^\frac{1}{4} r_\ast^2 \left[ \frac{\left(1 - \frac{\Lambda}{3} r^2_\infty \right)}{\sqrt{\frac{3m}{r_\ast} -1} + \sqrt{ \frac{3m}{r_\ast} - 1  + \frac{\Lambda r_\ast^4}{3r_\infty^2}}} \right]^\frac{3}{2}. \]

\subsubsection{Isothermal fluids: stiff equation of state}

In this Section we take the equation of state $p=e$. Equations (\ref{gb}) and (\ref{gc}) can be now written as
\[ r^2 \rho u^r = \mathrm{const}, \]
\[ \rho \sqrt{ 1 - \frac{2m}{r} - \frac{\Lambda}{3} r^2 + \left( u^r \right)^2 } = \mathrm{const}. \]
Thus
\[ (u^r)^2 = \frac{1 - \frac{2m}{r} - \frac{\Lambda}{3}r^2}{C r^4 - 1}, \]
where $C$ is some constant. The value of $C$ can be  fixed by requiring that the flow remains regular on the black hole horizon, i.e., for
\[ r = r_h = \left\{ \begin{array}{ll}
\frac{2}{\sqrt{\Lambda}} \cos \left[ \frac{\pi}{3} + \frac{1}{3} \mathrm{arc \, cos} \left( 3 m \sqrt{\Lambda} \right) \right], & 0 < \Lambda < \frac{1}{9m^2}, \\
2m, & \Lambda = 0, \\
\frac{2}{\sqrt{|\lambda|}} \sinh \left[ \frac{1}{3} \mathrm{ar \, sinh} \left( 3 m \sqrt{|\Lambda|} \right) \right], & \Lambda < 0
 \end{array} \right.  \]
Thus $C r_h^4 = 1$, and
\begin{equation}
\label{ai}
(u^r)^2 = \frac{1 - \frac{2m}{r} - \frac{\Lambda}{3}r^2}{\left( \frac{r}{r_h} \right)^4 - 1};
\end{equation}
the velocity remains finite at $r=r_h$. That in turn ensures that also 
  the density $\rho$ remains finite at the horizon of the black hole. It is equal to
\[ \rho = \rho_\infty \sqrt{\frac{ \left( 1 - \left( \frac{r_h}{r} \right)^4 \right) \left( 1 - \frac{2m}{r_\infty} - \frac{\Lambda}{3} r_\infty^2 \right) }{ \left( 1 - \left( \frac{r_h}{r_\infty} \right)^4 \right) \left( 1 - \frac{2m}{r} - \frac{\Lambda}{3} r^2 \right) } }. \]
For $\Lambda > 0$, the density given by the above formula diverges at the cosmological horizon. It is also possible to construct a solution that is regular there, but the regularity condition cannot be satisfied simultaneously at both horizons. In either case, the solution passes through a sonic point located precisely at this (event or cosmological) horizon at which the solution is regular.

The accretion rate  $\dot B$ can be now computed as
\begin{equation}
\label{zzz3}
\dot B = 4 \pi |u_\infty^r| r_\infty^2 \rho_\infty = 4 \pi \rho_\infty r_h^2 \sqrt{\frac{1 - \frac{2m}{r_\infty} - \frac{\Lambda}{3} r_\infty^2}{1 - \left( \frac{r_h}{r_\infty} \right)^4}}.
\end{equation}
Note that for $\Lambda = 0$ we have $r_h = 2m$ and
\[ \dot B = \frac{16 \pi \rho_\infty m^2}{\sqrt{\left(1 + \frac{2m}{r_\infty} \right) \left( 1 + \left(\frac{2m}{r_\infty}\right)^2 \right)}} \to 16 \pi m^2 \rho_\infty \]
for $r_\infty \to \infty$. This coincides with the result obtained in \cite{petrich} for the spherically symmetric case.

The mass accretion rate reads
\[ \dot m =  8 \pi r_h^{-2} \left( r_\infty^2 |u_\infty^r| \right)^2 e_\infty = 8 \pi r_h^2 \frac{1 - \frac{2m}{r_\infty} - \frac{\Lambda}{3} r_\infty^2}{1 - \left( \frac{r_h}{r_\infty} \right)^4} e_\infty. \]

For $r_\infty \gg m$ the above formulae can be approximated as
\begin{equation}
\label{zzz4}
\dot B \approx 4 \pi \rho_\infty r_h^2 \sqrt{1 - \frac{\Lambda}{3} r_\infty^2}
\end{equation}
and
\[ \dot m \approx 8 \pi r_h^2 \left(1 - \frac{\Lambda}{3} r_\infty^2 \right) e_\infty. \]

\subsection{Further discussion}

 Each of the three models in question possesses two transonic solutions. One of them is supersonic at infinity,  singular at the black hole horizon, regular at the cosmological horizon (if one does exist) and can extend beyond the cosmological horizon. The only sensible interpretation, as argued above,  is that  this solution (with positive coordinate velocity $u^r$) represents a wind --- outflowing gas --- and the bulk of gas is disconnected from the event horizon. The other solution is regular at the horizon and singular at a cosmological horizon (if that exists)  --- that should be interpreted, as pointed above --- as the accretion flow; it is clear that the physical accreting gas  must be comprised within  the cosmological horizon and it can penetrate the interior of a black hole.  

In contrast to the polytropic case, all investigated isothermal solutions can be defined globally for $\Lambda < 0$, i.e., one can set $r_\infty = \infty$. This resembles the classic situation occurring in the Schwarzschild case where the accretion cloud can also formally extend to infinity \cite{michel}. Note that for $\Lambda < 0$ the asymptotic behavior of the solutions is different from that known from the Schwarzschild case.

For all three solutions the mass accretion rate $\dot m$  decrease with the increase of the cosmological constant $\Lambda $ --- provided that remaining data are fixed --- and it vanishes at a critical value of $\Lambda $. The corresponding behaviour of $\dot m$ is depicted in Fig.~\ref{fig23} --- these graphs are plotted basing on the analytic formulae given in this section, with $m = 1$ and $r_\infty = 10^6$.

The dependence of the accretion rate $\dot B$ on $\Lambda $ is quite similar, and we decided not to present any additional plots in this paper. Let us point, however, an interesting exception --- the system with the ultra-hard equation of state $p = e$. In this case $\dot B$ achieves a maximum in the Schwarzschild--anti-de Sitter sector. This fact follows directly from Eq.~(\ref{zzz3}).

\section{Accretion in the ``Swiss cheese model'' of FLRW geometry}

\begin{figure}
\setlength{\tabcolsep}{0.02\textwidth}
\begin{tabular}{@{}p{0.48\textwidth}p{0.48\textwidth}@{}}
\includegraphics[height=0.45\textwidth]{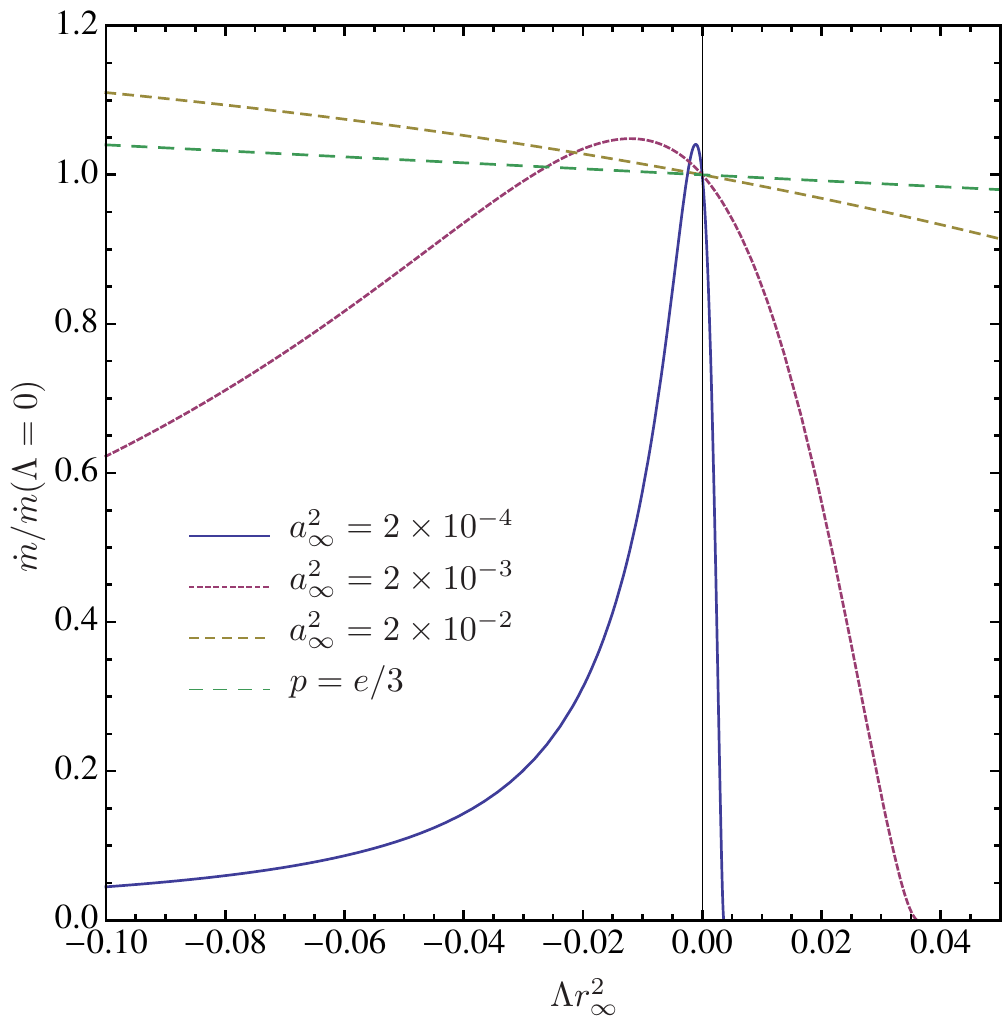}
&
\includegraphics[height=0.45\textwidth]{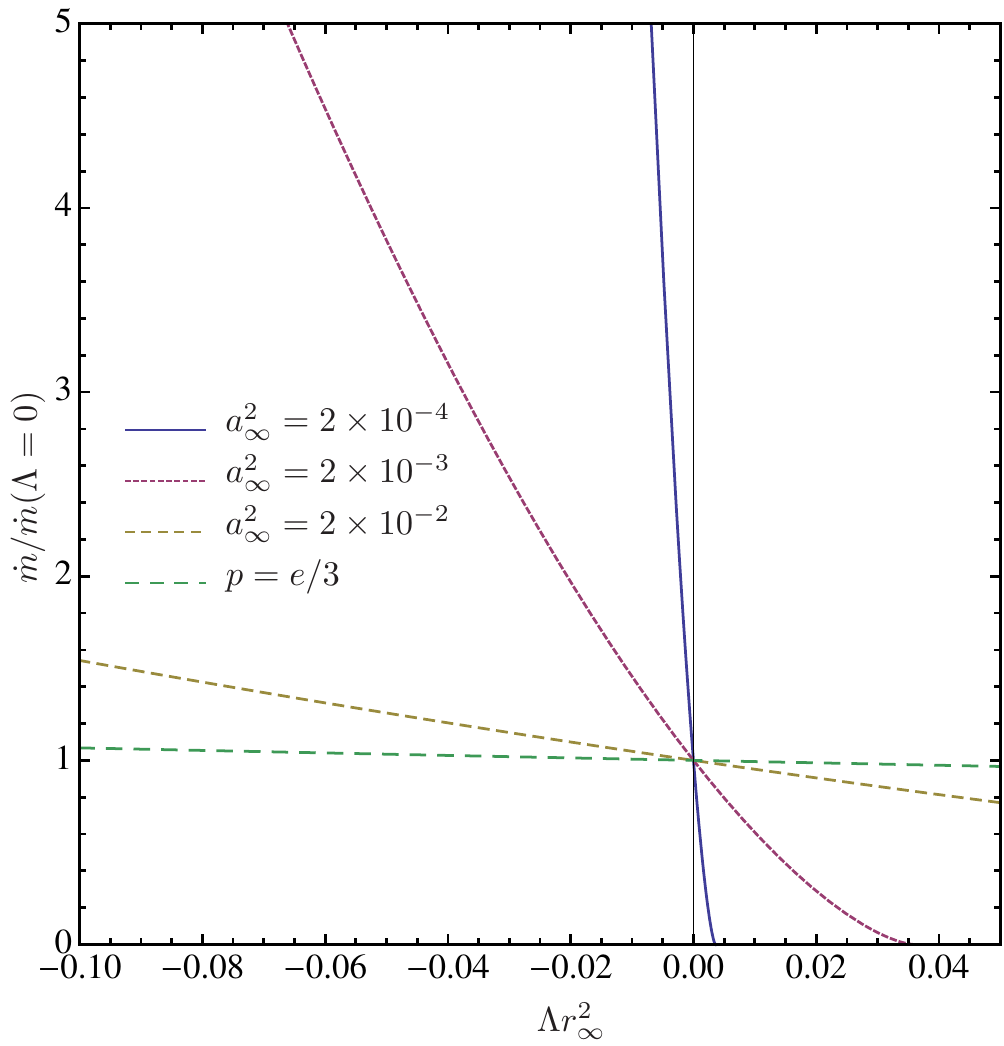}\\
\caption{\label{fig20} Dependence of the accretion rate $\dot m$ on $\Lambda$ for systems with fixed mass $m_f$. We plot data corresponding to polytropic fluids with $\Gamma = 4/3$, $m = 1$, $r_\infty = 10^6$ and $a_\infty^2 = 2 \times 10^{-4}, 2 \times 10^{-3}$, and $2 \times 10^{-2}$. The last graph depicts data obtained for the equation of state $p = e/3$.}
&
\caption{\label{fig21} Dependence of the accretion rate $\dot m$ on $\Lambda$ for system with fixed boundary energy density $e_\infty$. Here all parameters are exactly the same as in Fig.~\ref{fig20}.}
\end{tabular}
\end{figure}

\begin{figure}
\setlength{\tabcolsep}{0.02\textwidth}
\begin{tabular}{@{}p{0.48\textwidth}p{0.48\textwidth}@{}}
\includegraphics[height=0.45\textwidth]{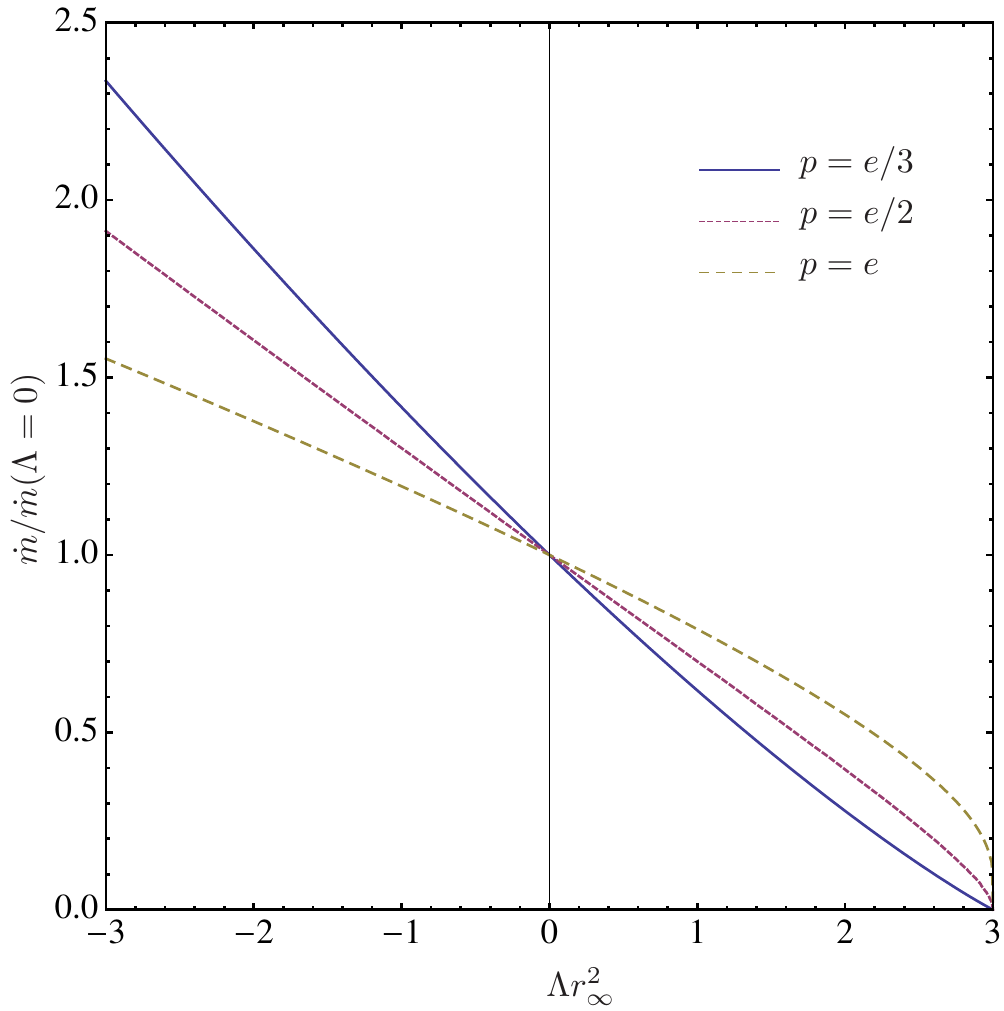}
&
\includegraphics[height=0.45\textwidth]{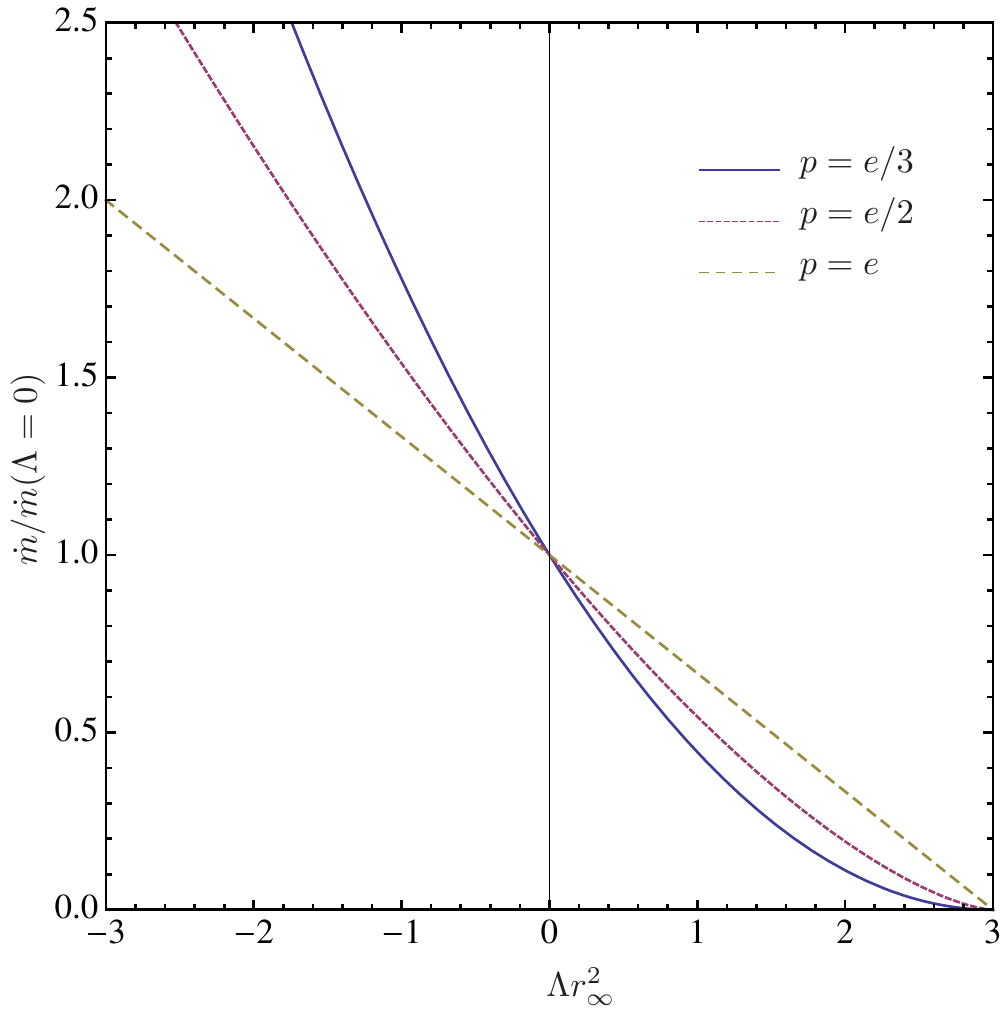}\\
\caption{\label{fig22} Dependence of the accretion rate $\dot m$ on $\Lambda$ for systems with fixed mass $m_f$. The three graphs correspond to equations of state $p = e/3$, $p = e/2$, and $p = e$. We assumed $m = 1$ and $r_\infty = 10^6$.}
&
\caption{\label{fig23} Dependence of the accretion rate $\dot m$ on $\Lambda$ for systems with fixed boundary energy density $e_\infty$. The three graphs correspond to equations of state $p = e/3$, $p = e/2$, and $p = e$. We assumed $m = 1$ and $r_\infty = 10^6$.}
\end{tabular}
\end{figure}

In this section we adopt our  accreting solutions to  a cosmological universe. The standard construction is the Einstein--Straus ``Swiss cheese'' model \cite{Einstein1}.  A spherically symmetric black hole surrounded with accreting gas is put inside a vacuole surrounded by a comoving boundary, that is matched (using the  Darmois--Israel junction conditions \cite{Darmois, Israel}) to the external FLRW spacetime. The cosmological constant permeates both the exterior and the interior of the vacuole, but outside of that only dust is allowed. The Darmois--Israel conditions demand that along the comoving boundary the first and second fundamental forms are continuous. In the spherically symmetric case the radius $R$ has to be continuous and in addition: i) the mass $m$ of the accreting system should be exactly equal to the mass of dust in the excised ball; ii) the boundary should be comoving with the Hubble velocity $H$. The whole gluing construction can be done explicitly. Balbinot et al.~have found an explicit solution that describes the corresponding matching between the internal Schwarzschild-de Sitter and external FLRW (dust + $\Lambda$) spacetimes \cite{Balbinot1988}. 

\subsection{Accretion rates inside a vacuole}

We wish to find how the presence of   a cosmological constant $\Lambda $ would affect  the   mass accretion rate $\dot m$ of a fluid onto the center.  It follows from the preceding discussion that when changing parameters of solutions while keeping isotropy and homogeneity of a cosmological geometry,  we must fix the total mass of the system comprised within the vacuole. Thus we shall investigate the dependence of the accretion rates on the cosmological constant, assuming the mass is fixed.

We define the mass of the fluid as
\[ m_f = 4 \pi \int_{r_h}^{r_\infty} dr r^2 e. \]
One can show --- using Eqs.~(\ref{ccab}) and (\ref{gb}) --- that for equations of state of the form $p = ke$ the mass of the fluid can be expressed as
\[ m_f = 4 \pi e_\infty (r_\infty^2 |u_\infty^r|)^{1+k} \int_{r_h}^{r_\infty} \frac{dr^\prime}{{r^\prime}^{2k} |u^r|^{1+k}}. \]
For $k = 1$ the above integral can be even evaluated analytically, but the resulting formulae are long and not very illuminating.

Expressing $e_\infty$ in terms of the corresponding $m_f$, and making use of Eq.~(\ref{nncg}), one obtains the following formula for $\dot m$:
\[ \dot m = \frac{1+k}{\sqrt{k} r_\ast^{2k}} \left( \frac{m}{2r_\ast} - \frac{\Lambda}{6} r_\ast^2 \right)^\frac{1-k}{2} \left( \int_{r_h}^{r_\infty} \frac{dr^\prime}{{r^\prime}^{2k} |u^r|^{1+k}} \right)^{-1} m_f. \]

An analogous result can be also obtained in the test-fluid polytropic case. A lengthy calculation allows to express the mass of the fluid as
\[ m_f = 4 \pi \rho_\infty \frac{\Gamma - 1}{\Gamma} \left( \frac{\Gamma - 1 - a_\infty^2}{a_\infty^2} \right)^\frac{1}{\Gamma - 1} \int_{r_h}^{r_\infty} dr r^2 \left( \frac{a^2}{\Gamma - 1 - a^2} \right)^\frac{\Gamma}{\Gamma - 1} \frac{\Gamma - a^2}{a^2}. \]
Similarily, expressing $\rho_\infty$ in terms of $m_f$ we can obtain the following formulae for the accretion rates
\begin{eqnarray*}
\dot m & = & \left( \frac{m}{2 r_\ast} - \frac{\Lambda}{6} r_\ast^2 \right) \frac{r_\ast^2}{a_\ast^3} \left( \frac{a_\ast^2}{\Gamma - 1 - a_\ast^2} \right)^\frac{\Gamma}{\Gamma - 1} \Gamma \left( \int_{r_h}^{r_\infty} dr r^2 \left( \frac{a^2}{\Gamma - 1 - a^2} \right)^\frac{\Gamma}{\Gamma - 1} \frac{\Gamma - a^2}{a^2} \right)^{-1} m_f, \\
\dot B & = & \frac{\Gamma}{\Gamma - 1} \sqrt{\frac{m}{2 r_\ast} - \frac{\Lambda}{6} r_\ast^2} r_\ast^2 \left( \frac{a_\ast^2}{\Gamma - 1 - a_\ast^2} \right)^\frac{1}{\Gamma - 1} \left( \int_{r_h}^{r_\infty} dr r^2 \left( \frac{a^2}{\Gamma - 1 - a^2} \right)^\frac{\Gamma}{\Gamma - 1} \frac{\Gamma - a^2}{a^2} \right)^{-1} m_f,
\end{eqnarray*}
where we used Eqs.~(\ref{bb}) and (\ref{nnbb}).

Figures \ref{fig20} and \ref{fig22} show the dependence of the accretion rates $\dot m$ on the cosmological constant for systems with fixed masses. In both plots the accretion rates are normalized to unity for $\Lambda = 0$, so the actual value of $m_f$ is irrelevant. A particularily interesting situation occurs for polytropic fluids. This case is depicted in Fig.~\ref{fig20}, assuming $m = 1$, $\Gamma = 4/3$, $r_\infty = 10^6$, and $a_\infty^2 = 2 \times 10^{-4}, 2 \times 10^{-3}$, and $2 \times 10^{-2}$, respectively. For all values of $a_\infty$, the accretion rate tends to zero at some finite positive and some finite negative value of $\Lambda$, for which stationary solutions cease to exist (zeros corresponding to negative values of $\Lambda$ are not visible, due to limitations of the plot). The maximum of the accretion rate occurs at some small negative value of $\Lambda$, that depends on $a_\infty$. Figure \ref{fig20} also shows the accretion rate $\dot m$ for the constant mass system with the equation of state $p = e/3$. It is natural to expect that this should constitute a limit of a sequence of polytropes with $\Gamma = 4/3$ and increasing internal energy (sound speed) (cf., e.g.,~\cite{mach_pietka}). It is somewhat surprising to find out that for $p = e/3$ the dependence of the accretion rate on $\Lambda$ is strictly decreasing, and there is no zero in the sector of negative cosmological constants (the limit is probably still achieved, but in a point-wise sense). In this respect isothermal solutions differ qualitatively from the polytropic ones.

Figure \ref{fig22} shows analogous relations for systems with equations of state $p = e/3$, $p = e/2$, $p = e$, and fixed masses of the fluid. In this case we also assumed $m = 1$ and $r_\infty = 10^6$. The obtained relations are strictly monotonic --- all accretion rates decrease with $\Lambda$. A zero is achieved at $\Lambda r_\infty^2 = 3 \left( 1 - 2 m /r_\infty \right)$, i.e., when the location of the cosmological horizon coincides with the assumed outer boundary of the accretion cloud.

Figures \ref{fig21} and \ref{fig23} show the dependence of $\dot m$ on $\Lambda$ for systems with fixed boundary values of the densities $\rho_\infty$ or $e_\infty$. Figure \ref{fig21} presents results for polytropes with the same parameters as in Fig.~\ref{fig20}. Quite surprisingly, in all examined cases the dependence of $\dot m$ on $\Lambda$ appeared to be strictly decreasing.  

The fact that the relation between the accretion rate and the cosmological constant depends, for polytropic fluids, so strongly on the parametrization of the solutions is connected with the behavior of the density for $\Lambda <0$, and ultimately with the existence of the ``homoclinic-type'' solutions. In the standard Schwarzschild case \cite{michel} with realistic parameters of the model, the density of the fluid is nearly constant in the bulk of the accretion cloud, so the total mass of the fluid is practically proportional to the boundary density \cite{kkmms}. Thus both parametrizations --- fixing the total mass of the accretion cloud, or the boundary value of the density --- are basically equivalent. This is no longer true for polytropic fluids and $\Lambda < 0$, in which case the density varies considerably for large radii --- this can be seen in Fig.~\ref{figb} (note that $\rho \sim 1/(r^2 u^r)$).

\subsection{Self-gravitating polytropic flows}
 
Below we shall include self-gravity of fluids and find numerically transonic accretion flows. The relevant equations and the details of the numerical procedure can be found in \cite{karkowski_malec_2013}. It is convenient to work in the comoving coordinates. The line element reads
\[ ds^2 = - N^2 dt^2 + \tilde a d r^2 + R^2  \left( d \theta^2 + \sin^2 \theta d \phi^2 \right), \]
where $N$ (the lapse), $\tilde a$ and the areal radius $R$ are functions of $t$ and $r$. We assume the total mass $m = 1$, $\Gamma =4/3$, $R_\infty =10^6$, and $a_\infty^2 = 2 \times 10^{-4}$. The boundary values of the unknown metric function $N$ and the mean curvature of centered two-spheres $k$ of constant $R$ at $R_\infty$ are
\[ k = N =\sqrt{1 -  \frac{2m}{R_\infty} - \frac{\Lambda}{3} R^2_\infty + U_\infty^2 }, \]
where $U = N^{-1} dR/dt$. The functions $N$, $k$ and the velocity $U$ completely specify the spacetime geometry. There are free parameters: the cosmological constant $\Lambda$, the mass accretion rate $\dot B$ and the asymptotic density $\rho_\infty$ --- after specifying them, equations are integrated inward, starting from $R_\infty$. These quantities have to satisfy the asymptotic conditions: $U_\infty^2 \ll 2m/R_\infty \ll  a^2_\infty$.

The calculations had been performed for a few dozens of negative values of the cosmological constant $\Lambda $. We take $\Lambda R_\infty^2 \in \left( -3 \times 10^{-3}, 0 \right)$. For each fixed value of $\Lambda$ we find solutions corresponding to different values of $\dot m$.

\begin{figure}
\begin{center}
\includegraphics[height=0.45\textwidth]{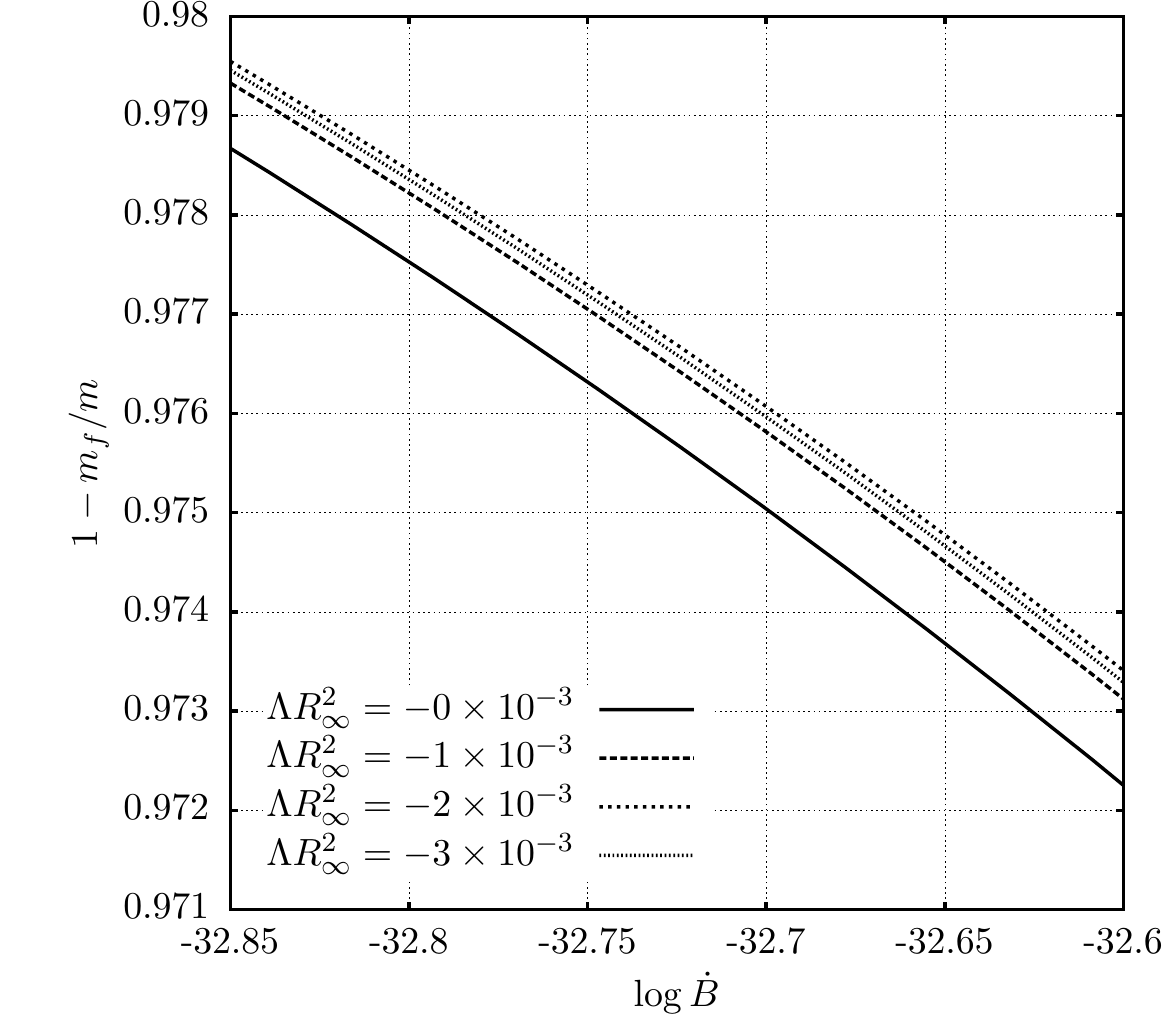}
\end{center}
\caption{\label{fig:1} The abscissa shows the logarithm of mass accretion rate  $\dot m$ and the ordinate shows $1 - m_f/m$ for selfgravitating polytropes with $\Gamma = 4/3$, $R_\infty=10^6$, $a_\infty^2 = 2 \times 10^{-4}$ and the total mass $m = 1$. The  various lines  correspond  to $\Lambda R^2_\infty/10^{-3} = 0, -1, -2, -3$. The maximum of $\dot m $ is achieved around $\Lambda R_\infty^2 = -2$.}
\end{figure}

Fig.~\ref{fig:1} displays obtained results in the regime of test gases --- its fractional contribution varies between 2--3\% of the total mass. The ordinate shows the ratio $1 - m_f/m$ of the mass
\[ m_f = 4\pi \int_{R_h}^{R_\infty }dr r^2 e \]
of gas to the total mass $m$. Here, as before, $R_h$ denotes the areal radius of the horizon of the central black hole. The total mass has been normalized to 1. The abscissa shows $\log \dot B$.  It is clear that the mass accretion rate increases with the increase of $|\Lambda|$ for $\Lambda R_\infty^2  \in \left( -2 \times 10^{-3}, 0 \right)$ --- that feature was not discovered in \cite{karkowski_malec_2013} --- and then starts to decrease. Similar results can be obtained for systems with higher gas content, with analogous conclusions. This behaviour agrees with what we found earlier in the case of test polytropes.
 
It follows from the inspection of the figure that the sensitivity of the mass accretion rate onto the cosmological constant is quite prononuced --- $\dot B$ increases by a few percents between $\Lambda R_\infty^2 = 0$ and $\Lambda R_\infty^2 = -2\times 10^{-3}$.
 
\section{Conclusions}

We have found new exact solutions that describe the   spherical steady accretion of isothermal test fluids ($p=ke$, $k=1/3, 1/2, 1$) in Schwarzschild--(anti-)de Sitter spacetimes. We discovered numerically  the  existence of ``homoclinic'' solutions --- solutions for which the graph of the square of the radial velocity versus the areal radius forms a closed loop joining the same critical point. It is interesting that they  exist only for polytropic equations of state and for a negative cosmological constant.
 
We found that when keeping fixed the asymptotic mass density of fluids, the mass accretion rate decreases with the increase of $\Lambda$. This is true for polytropic and isothermal fluids, with the exception of the stiff equation of state  $p = e$. In this particular case the baryonic mass accretion rate achieves a maximum at some $\Lambda < 0$ (it is still a decreasing function of $\Lambda$ for $\Lambda > 0$). On the other hand, when the mass of  the accreting system is kept constant, then polytropic and isothermal fluids behave in a different way under the change of $\Lambda$. Polytropic accreting flows achieve a maximum mass accretion rate at a negative value of $\Lambda$. The investigated isothermal fluids accrete with the efficiency that monotonically decreases with the increase of $\Lambda$. These constatations are relevant cosmologically.

The interesting feature of polytropic accretion flows is that the accretion stops completely when the \textit{absolute value} of $\Lambda$ is large enough. The accretion flows stop for both isothermal and polytropic fluids when \textit{positive} $\Lambda$ is large enough. This agrees with earlier findings valid only for polytropic fluids \cite{karkowski_malec_2013}.

These results shed light on the possible influence of the cosmological constant $\Lambda$ onto the formation of material structures. They allow one to expect, that if the cosmological constant enhances/damps Bondi accretion, then realistic models should reveal more/less robust structure formation.  Moreover, the mere fact that the robustness of accretion is $\Lambda$-dependent opens, at least in principle, a way to estimate the value of the cosmological constant by a quasilocal analysis.
 
The ``Weyl curvature hypothesis'' \cite{Penrose2004} asserts, in its informal version, that a Friedman--Lemaitre--Robertson--Walker spacetime, where the Ricci curvature is nonzero but the Weyl curvature vanishes, evolves towards a vacuum spacetime filled with black holes and gravitational radiation ---  with nonvanishing Weyl curvature and  negligible Ricci tensor. This scenario assumes that initially small inhomogeneities of FLRW universe accrete matter and transform themselves into black holes, that gradually merge, leaving at the end a net of huge  black holes and a lot of gravitational radiation. Our findings suggest that the role of steady accretion in the realization of this scenario, in the presence of dark energy, is insignificant.  This can be interpreted as indication that the cyclic universe scenario of Penrose \cite{Penrose2004} cannot work.

\section*{Acknowledgments}

This work has been partially supported by the following grants: Polish Ministry of Science and Higher Education grants IP2012~000172 (Iuventus Plus), 7150/E-338/M/2013, and the NCN grant DEC-2012/06/A/ST2/00397.

\end{document}